\documentclass[twocolumn,times]{aastex62}
\usepackage{xcolor, bm, tablefootnote, amsmath}

\newcommand{\uw}{University of Washington Astronomy Department, Box 351580, Seattle, WA 98195-1580, USA}
\newcommand{\hi}{$\mathrm{H} \, \textsc{i} \,$}

\newcommand{\sigmadust}{$\Sigma_\mathrm{dust}$\,}

\newcommand{\sigmametalgas}{$\Sigma_\mathrm{metal}^\mathrm{gas}$\,}
\newcommand{\sigmametalstar}{$\Sigma_\mathrm{metal}^\star$\,}
\newcommand{\sigmametaldust}{$\Sigma_\mathrm{metal}^\mathrm{dust}$\,}
\newcommand{\sigmaprod}{$\Sigma_\mathrm{metal}^\mathrm{produced}$\,}
\newcommand{\sigmapresent}{$\Sigma_\mathrm{metal}^\mathrm{present}$\,}

\newcommand{\xco}{$X_\mathrm{CO}$\,}

\shorttitle{Metal Loss from M31}
\shortauthors{Telford et al.}

\begin{document}
\correspondingauthor{Grace Telford}
\email{otelford@uw.edu}

\title{Spatially Resolved Metal Loss from M31}
\author[0000-0003-4122-7749]{O. Grace Telford\footnotetext{}}
\altaffiliation{NSF Graduate Research Fellow}
\affiliation{\uw}
\author[0000-0002-0355-0134]{Jessica K. Werk}
\affiliation{\uw}
\author[0000-0002-1264-2006]{Julianne J. Dalcanton}
\affiliation{\uw}
\author[0000-0002-7502-0597]{Benjamin F. Williams}
\affiliation{\uw}

\begin{abstract}

As galaxies evolve, they must enrich and exchange gas with the surrounding medium, but the timing of these processes and how much gas is involved remain poorly understood. In this work, we leverage metals as tracers of past gas flows to constrain the history of metal ejection and redistribution in M31. This roughly $L*$ galaxy is a unique case where spatially resolved measurements of the gas-phase and stellar metallicity, dust extinction, and neutral interstellar gas content are all available, enabling a census of the current metal mass. We combine spatially resolved star formation histories from the Panchromatic Hubble Andromeda Treasury survey with a metal production model to calculate the history of metal production in M31. We find that $1.8\times10^9 \, M_\odot$ of metals, or 62\% of the metal mass formed within $r < 19 \,\mathrm{kpc}$, is missing from the disk in our fiducial model, implying that the M31 disk has experienced significant gaseous outflows over its lifetime. Under a conservative range of model assumptions, we find that between 3\% and 88\% of metals have been lost ($1.9\times10^7 - 6.4\times10^9 \, M_\odot$), which means that metals are missing even when all model parameters are chosen to favor metal retention. We show that the missing metal mass could be harbored in the circumgalactic medium of M31 if the majority of the metals reside in a hot gas phase. Finally, we find that some metal mass produced in the past 1.5 Gyr in the central $\sim5 \,\mathrm{kpc}$ has likely been redistributed to larger radii within the disk.

\end{abstract}


\section{Introduction\label{sec:intro}}

\subsection{Metals as Tracers of Past Gas Outflows\label{sec:background}}

Galaxy formation models require gaseous outflows to regulate star formation and reproduce the observed scaling relations, such as the the star-forming main sequence (SFMS) and the mass-metallicity relation \citep[e.g.,][]{somerville15}. To explain the mass-metallicity relation in particular, a larger fraction of metals must be ejected preferentially from lower-mass galaxies \citep[e.g.,][]{tremonti04, dalcanton07, peeples11}. These ejected metals would also explain the observed enrichment of the circumgalactic medium (CGM) and intergalactic medium \citep[IGM; e.g.,][]{oppenheimer06, werk14}.

Observations have shown that outflows are ubiquitous at high redshift \citep[e.g.,][]{shapley03, weiner09, steidel10} and occur locally in starburst and post-starburst galaxies \citep[e.g.,][]{tremonti07, mcquinn-k10, chisholm18}. Although it is now widely accepted that outflows drive baryons and metals out of essentially all galaxies at some point in their evolution, how these outflows are launched and the fate of the ejected material remain poorly understood, which is due in large part to the difficulty in characterizing diffuse and multiphase outflows for large numbers of galaxies. The properties of outflows are also expected to vary strongly in time, which further complicates the interpretation of instantaneous measurements of outflowing material.

Attempts to use theoretical models to interpret observational constraints on outflows have had mixed results. Many different feedback implementations in galaxy formation models are able to reproduce key properties of the galaxy population \citep[e.g.,][]{naab17}, but predict different amounts of metal loss from galaxies \citep[e.g.,][]{wiersma11}. Recent particle-tracking analyses do not agree on the fractions of baryons and metals in various reservoirs at low redshift: stars, cold interstellar medium (ISM) gas, CGM, and IGM \citep{ford14, christensen16, christensen18, angles-alcazar17}. Furthermore, these studies predict different scalings for the amount of metal loss with galaxy mass, as well as different timescales on which previously ejected metal-enriched material is reaccreted. In light of these discrepancies, observational constraints on the metal mass present in these various reservoirs represent a promising route toward distinguishing among these various feedback models.

Several observational studies have calculated the global fraction of metal mass that is retained in galaxies across a wide stellar mass range \citep[e.g.,][]{zahid12b, peeples14}. These studies use scaling relations to estimate metal retention, and so represent an average constraint; it remains unclear how variable the net metal loss from galaxies of similar stellar mass might be. While it is known that dwarf galaxies can lose most of their metal mass \citep{kirby11, mcquinn-k15}, the scaling of metal retention with galaxy mass for high-mass galaxies is not yet settled.

Spatially resolving the baryon content and metallicities enables a more precise calculation of the total metal mass present in a galaxy. Both stellar and gas-phase metallicity gradients are known to be common \citep[e.g.,][]{sanchez13, roig15}, so applying a metallicity measured in a galaxy center to its entire stellar or gas content is likely to bias results. Only one such spatially resolved measurement of metal retention within a galaxy has been reported to date, which is for NGC 628 \citep{belfiore16a}. 

In this work, we leverage the wealth of data available for the nearby galaxy M31 to perform a spatially resolved measurement of its lifetime metal retention. This is the first detailed measurement for an $L*$ galaxy, providing an important anchor for population-wide studies and comparisons to galaxy formation models.

\subsection{This Work: A Spatially Resolved Measurement of Metals Missing from M31\label{sec:roadmap}}

\begin{figure*}[!ht]
\minipage{0.65\textwidth}
  \includegraphics[width=\linewidth]{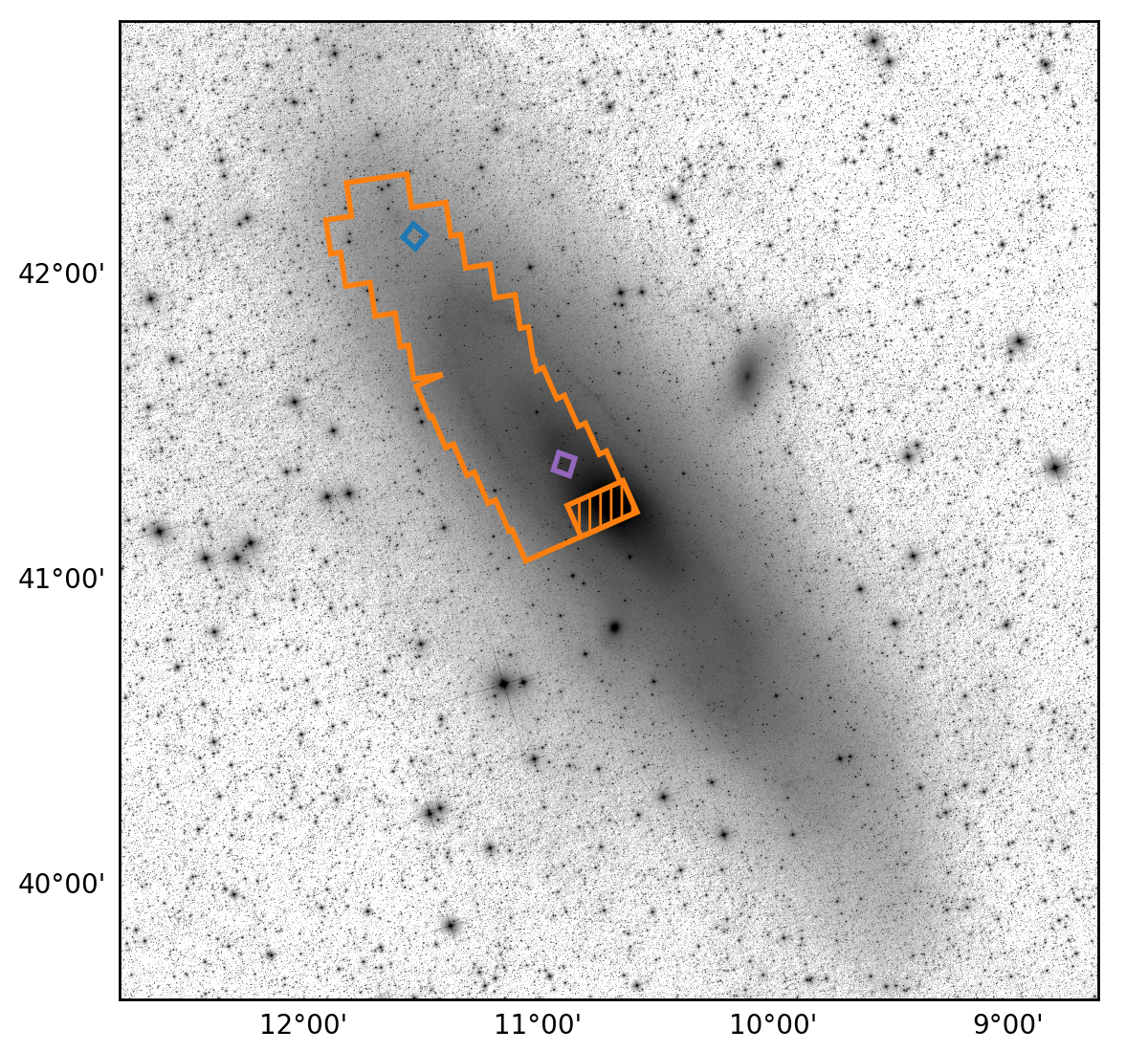}
\endminipage\hfill
\minipage{0.35\textwidth}
  \includegraphics[width=\linewidth]{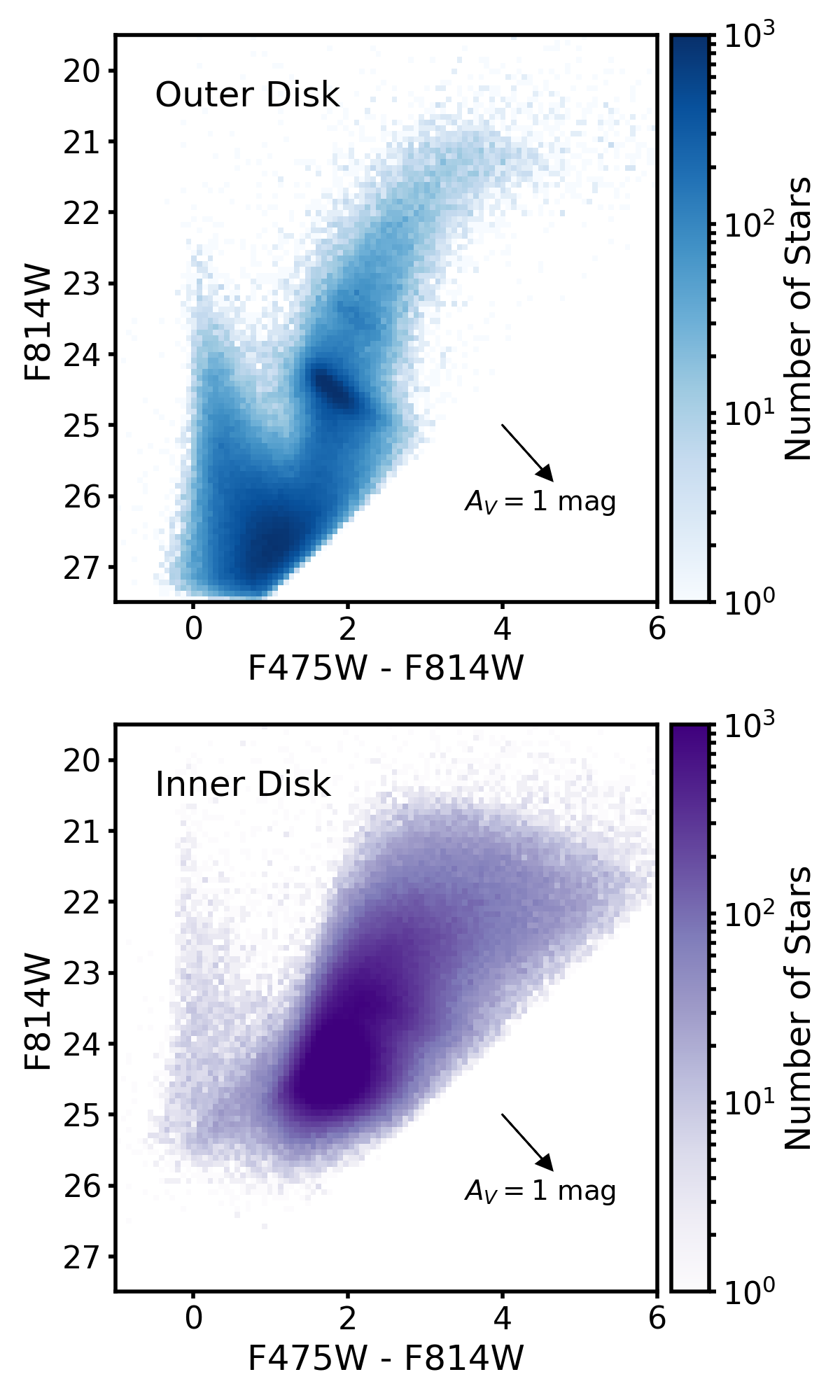}
\endminipage\hfill
\caption{\textsc{\textbf{Overview of the PHAT survey and stellar photometry data.}} Left: the footprint of the PHAT survey overplotted on a 3.4 $\mu$m image of M31 \citep{lang14}. The orange outline shows the area covered by the survey, and the hatched rectangle shows the region covering the bulge that was excluded from the \citet{williams17} SFH analysis due to crowding. The blue and purple squares show the regions containing the stars in the CMDs in the right panel. Right: example optical CMDs from the outer (top) and inner (bottom) regions of M31's disk. Darker colors indicate more densely populated regions of the CMDs, and the arrows illustrate the effect of 1 mag of dust extinction in the $V$ band on a star's position in the CMD. Stellar crowding limits photometric depth, resulting in a shallower CMD for the inner disk region. The total SFH shown in Figure~\ref{fig:sf_z_history} below was inferred from modeling the distribution of stars in CMDs across the PHAT footprint (described in Section~\ref{sec:sfhs}).
\label{fig:phat_overview}}
\end{figure*}

\begin{figure*}[!ht]
  \includegraphics[width=\linewidth]{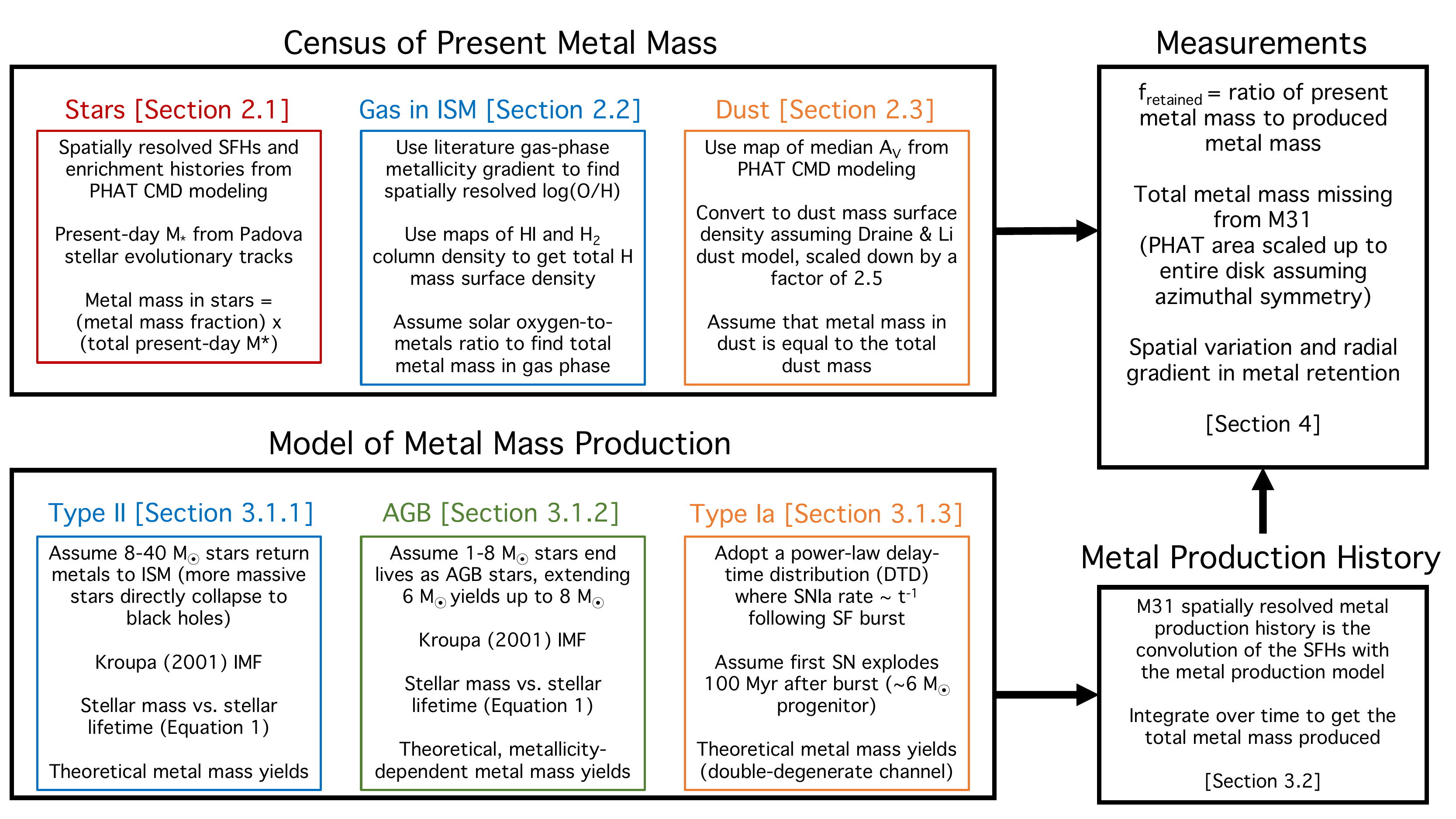}
\caption{\textsc{\textbf{Summary of methods for calculating present and produced metal mass.}} This visual aid shows the data and model ingredients used in our calculations of the spatial distribution of metals currently in M31 (Section~\ref{sec:metals_present_all}) and the spatially resolved metal production history (Section~\ref{sec:metal_history_all}). Colors correspond to those used in later plots to represent each metal reservoir (Figure~\ref{fig:metals_present}) or nucleosynthetic source (Figures~\ref{fig:metal_model} and~\ref{fig:metal_history}). Ultimately, we integrate the metal production histories over time to map the total produced metal mass, and divide by the total present metal mass calculate the metal retention fraction, $f_\mathrm{retained}$ (Section~\ref{sec:results}).
\label{fig:flowchart}}
\end{figure*}

M31 is the nearest massive ($\sim L*$) galaxy \citep[at a distance of 785 kpc,][]{mcconnachie05}. Because of its brightness, large area on the sky, and similarity to the Milky Way, M31 is extremely well studied. In particular, photometry of individual resolved stars is available from the \textit{Hubble Space Telescope (HST)} imaging in the Panchromatic Hubble Andromeda Treasury \citep[PHAT;][]{dalcanton12, williams14}. PHAT enabled precise photometric measurements for over 100 million stars in six filters spanning the UV to IR \citep{williams14}. The survey covers roughly one-third of M31, particularly the near side of the northern disk out to $\sim 20$ kpc. The \textit{HST} imaging is complemented by spatially resolved maps of the cold gas (\hi and H$_2$), which we describe in Section~\ref{sec:gas_maps} below.

The PHAT photometry has enabled measurements of spatially and temporally resolved star formation histories (SFHs), enrichment histories, and dust content derived from modeling color-magnitude diagrams \citep[CMDs;][described in Sections~\ref{sec:sfhs} and ~\ref{sec:dustmap} below]{williams17, dalcanton15}. M31 is the most massive nearby galaxy for which these precise CMD-based measurements are possible. The left panel of Figure~\ref{fig:phat_overview} shows the footprint of the PHAT survey as the orange outline on a $3.4\,\mu\mathrm{m}$ image of M31 from the \textit{Wide field Infrared Survey Explorer} \citep{wright10, lang14}. The hatched region over the bulge of the disk was excluded from the \citet{williams17} SFH analysis because stellar crowding in the densest regions limits the depth of the CMDs and therefore the reliability of SFH determinations. The blue and purple squares show the areas for which example optical CMDs are shown in the upper and lower right panels, respectively. Darker colors in the CMDs indicate that more stars populate these regions. The CMD for stars in the outer disk (upper panel, blue) reaches fainter magnitudes than that for the inner disk (lower panel, purple) because crowding limits photometric quality in the inner disk. The larger spread in the red giant branch (RGB) in the inner disk CMD is due to a larger age and metallicity spread there.

In this work, our main objective is to measure the lifetime metal mass loss from M31. An overview of the measurement is shown in Figure~\ref{fig:flowchart}, which illustrates the datasets and model assumptions that enter into each step. We first perform a census of the spatial distribution of metals in the stars, ISM gas, and dust within the PHAT footprint in the M31 disk (Section~\ref{sec:metals_present_all}). We then calculate the spatially and temporally resolved history of metal production in M31 (Section~\ref{sec:metal_history_all}), and from this, the mass surface density of metals produced by stars currently in the galaxy, \sigmaprod. We take ratio of the present-day metal mass surface density, \sigmapresent, to \sigmaprod to derive both the integrated and spatially resolved metal retention fraction, $f_\mathrm{retained}$ (Section~\ref{sec:results}).

Throughout this calculation, we note the dominant sources of systematic uncertainty and place conservative bounds on the possible values of each quantity of interest. We do not fully model the effects of stellar radial migration, merger-driven accretion of stars and/or pre-enriched gas, or recycling of metal-enriched winds within the galaxy, but we discuss the effects of these processes on our results in Section~\ref{sec:results}. 

Finally, we place constraints on the lifetime-averaged mass loading of outflows, the metal content of the M31 CGM, and on the required spatial redistribution of metals recently produced in the inner disk in Section~\ref{sec:discussion}. Our findings are summarized in Section~\ref{sec:conclusions}.

We adopt a flat $\Lambda\mathrm{CDM}$ cosmology with $\Omega_{m}=0.308$ and $H_0=67.8 \,\mathrm{km\,s}^{-1}\mathrm{Mpc}^{-1}$ \citep{planck16-cosmo}. We use $\Sigma$ to refer to mass surface density throughout. $Z$ is the metal mass fraction, including all elements heavier than He. The mass fraction of an individual metal species, e.g., oxygen, is referred to as $Z(\mathrm{O)}$. We use the \citet{anders89} solar abundance set with $Z_\odot=0.019$ (which is consistent with the abundance set used in the Padova stellar evolutionary tracks, as discussed in Section~\ref{sec:sfhs} below). We also quote logarithmic stellar metallicities relative to solar, $\mathrm{[M/H]} = \log (Z/Z_\odot)$.


\section{The Spatial Distribution of Metals Present in M31\label{sec:metals_present_all}}

Here, we calculate the metal mass surface density that is present in the PHAT footprint, and the implied total metal mass in M31 assuming azimuthal symmetry. We describe the data sources and methods of calculating the metal mass surface density in stars (\sigmametalstar), cold ISM gas (\sigmametalgas), and dust (\sigmametaldust), as well as the sources of systematic uncertainty in each measurement. We present the radial profiles of \sigmametalstar, \sigmametalgas, and \sigmametaldust and calculate the fractional contribution of each reservoir to the total metal content.

\begin{figure*}[!ht]
\minipage{0.5\textwidth}
  \includegraphics[width=\linewidth]{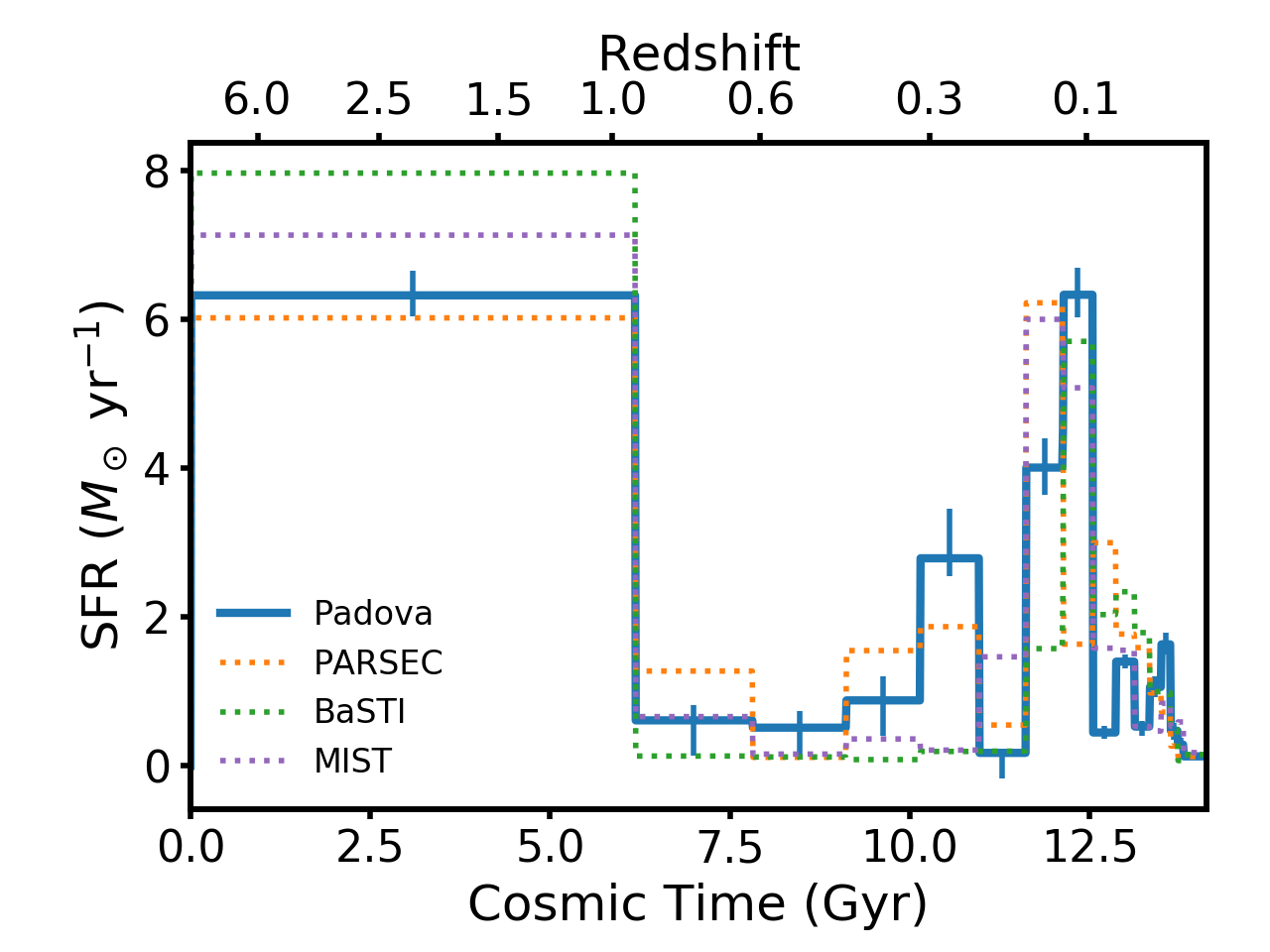}
\endminipage\hfill
\minipage{0.5\textwidth}
  \includegraphics[width=\linewidth]{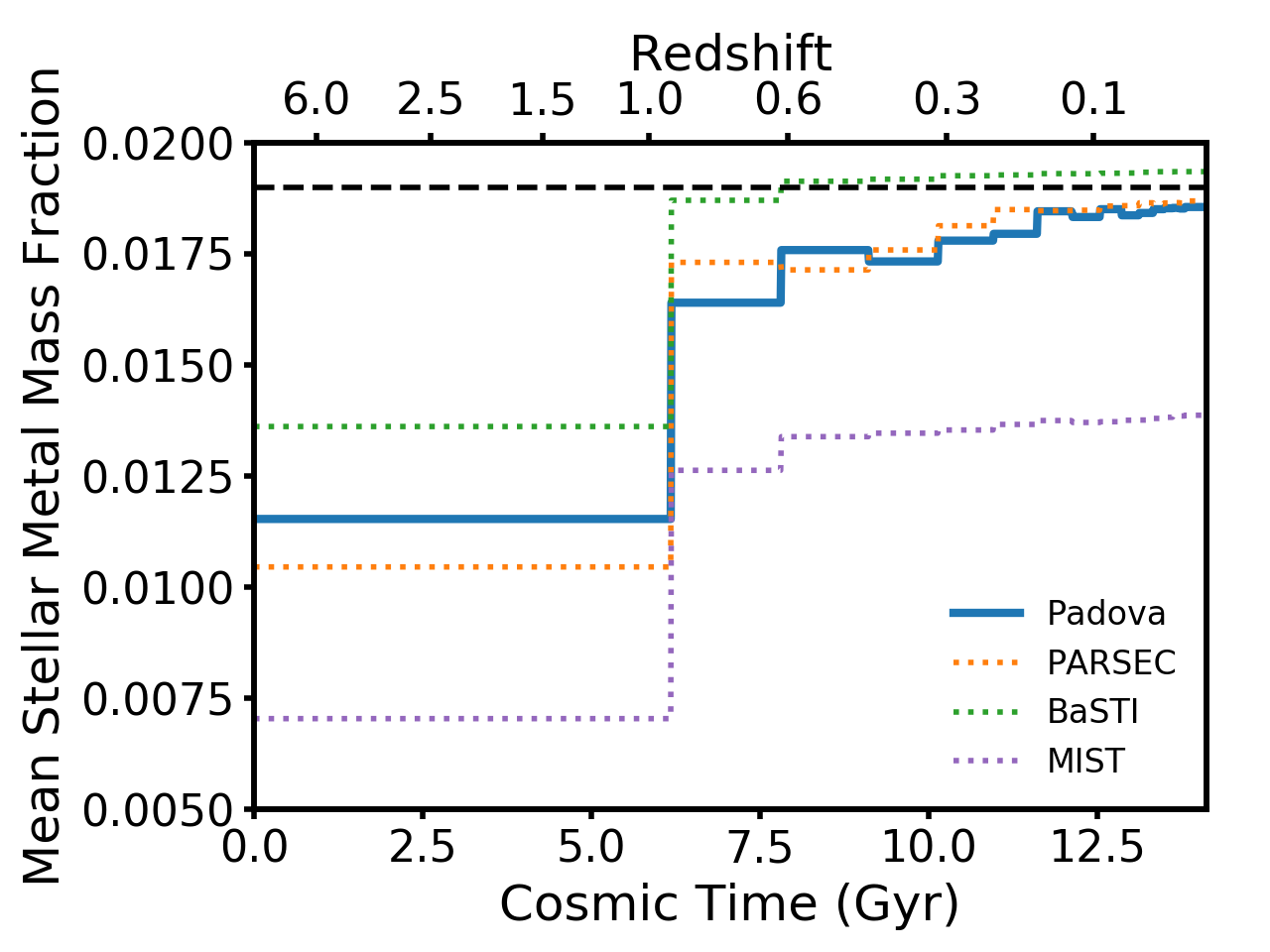}
\endminipage\hfill
\caption{\textsc{\textbf{The star formation and stellar enrichment history in M31 from \citet{williams17}.}} Left: total star formation rate (SFR) within the PHAT footprint plotted against the age of the universe (with redshift shown on the top axis for reference). Right: mean metal mass fraction of newly formed stars in each age bin plotted against time. The bulk of star formation occurs during the oldest bin ($z\gtrsim1$), which is wider than the other age bins because the data cannot constrain variations in SFR at ancient times. In both panels, the solid blue line shows the fiducial model, derived using Padova stellar evolutionary tracks. The error bars on the Padova SFRs in the left panel show the random uncertainties in the fit. The orange, green, and purple dotted lines show results for different stellar tracks that are used; the spread among these is used to gauge the systematic uncertainty in the SFHs and stellar metallicities. In the right panel, the horizontal black dashed line indicates solar metallicity ($Z_\odot=0.019$), which is adopted by the Padova models. 
\label{fig:sf_z_history}}
\end{figure*}

\subsection{Stars\label{sec:stars}}

\subsubsection{Ancient Star Formation and Enrichment Histories\label{sec:sfhs}}

The ancient SFHs were measured by \citet{williams17}, who modeled optical to near-infrared (NIR) CMDs within 826 regions $83''\times83''$ in size, corresponding to a physical size of $0.3\times1.4$ kpc \citep[corrected for the 77$^{\circ}$ inclination of M31, e.g.,][]{courteau11}. We call these regions ``SFH pixels'' throughout the paper. We summarize here the key modeling choices made by \citet{williams17}, but refer the reader to that paper for details.

\citet{williams17} constructed CMDs using photometry in the Advanced Camera for Surveys (ACS) F475W and F814W optical bands and in the Wide Field Camera 3 (WFC3) F110W and F160W NIR bands. They modeled these optical-NIR CMDs using \textsc{match} \citep{dolphin02, dolphin12, dolphin13}. They used a logarithmic age grid with 0.1 dex resolution from $\log(t/\mathrm{yr}) = 8.5-9.9$, and wider time bins at the oldest and youngest ages: $\log(t/\mathrm{yr})=6.6-8.5$ and $\log(t/\mathrm{yr})={9.9-10.15}$. They adopted a  \citet{kroupa01} initial mass function (IMF) for primary stars, and 30\% of these were assumed to have a binary companion. 

Dust affects the features in the CMDs that provide constraints on the SFH and metallicity. \citet{williams17} incorporated prior knowledge of the dust distribution from the \citet{dalcanton15} dust maps (described in Section~\ref{sec:dustmap}) in their modeling to distinguish the effects of dust and stellar population properties on the CMD features. Briefly, they adopted a lognormal $A_V$ distribution model in each SFH pixel and fixed the lognormal parameters based on the best-fit parameters from the \citet{dalcanton15} maps (see \citealt{williams17} for a more detailed explanation). They also included a uniform foreground dust screen in the dust model, which was a free parameter that was independently fit in each SFH pixel.

The oldest main-sequence turnoffs are not resolved in the PHAT CMDs because stellar crowding prevents reliable detection of fainter stars (see the right panels in Figure~\ref{fig:phat_overview}). This limits the ability of \textsc{match} to simultaneously constrain the stellar metallicity and age, such that when both variables are allowed to be free in the fitting, unphysical variations in stellar metallicity at a given age between adjacent SFH pixels can occur. To ensure that the stellar metallicities vary smoothly in space and in time, \citet{williams17} fixed age-metallicity relations (i.e., they enforced a mean [M/H] at each age in each SFH pixel), adopting the exponentially declining enrichment rates that provided the best fits to the data in three radial bins such that inner regions enrich earlier. 

The stars formed in each age bin are still allowed span a range of metallicities, with 0.25 dex spread in [M/H] (see Figure 8 in \citealt{williams17}). Given the adopted age-metallicity relation and dust model, the best-fit SFH is that which best reproduces the observed number of stars in different regions of the CMD (nominally, the main sequence, asymptotic giant branch (AGB), and He-burning sequences at ages younger than about 2 Gyr, and the red giant branch (RGB) and red clump at older ages). For each SFH pixel, the output of the CMD modeling is an SFH, where each fitted epoch of time is broken down into 24 bins of [M/H], or equivalently, $Z = Z_\odot \times10^\mathrm{[M/H]}$, in each age bin. The total SFR in a given age bin is therefore the sum of the SFRs in all metallicity bins at that age. In each age bin, we know the SFR (and therefore stellar mass formed) in all metallicity bins, we can calculate the mean stellar metal mass fraction as the mass-weighted mean $Z$ of stars that formed in a given age bin.

The adopted age-metallicity relations are physically motivated and likely provide a more realistic enrichment history than would be obtained if metallicity were a free parameter in the CMD modeling. Still, we have checked that our quantitative results and conclusions are only affected at the few-percent level if we instead adopt the SFHs and enrichment histories that would be recovered when the metallicities in each age bin were free parameters in the CMD modeling. \citet{williams17} showed that the enrichment histories are difficult to reliably constrain with photometry of the depth and quality of PHAT, showing rapid changes at intermediate and recent times when model CMD features are less sensitive to metallicity across all model sets (see their Figure 20). However, the best-fit metallicity at the earliest times is consistent with the adopted age-metallicity relations, and our calculations are more sensitive to the early enrichment because most stellar mass formed in the oldest age bin.

Finally, we note that the lack of constraints on the population of low-mass stars in the CMDs results in a systematic uncertainty in the normalization of the SFHs and total stellar mass formed. This uncertainty at the factor of $\sim$2 level is known to affect most stellar mass measurement techniques \citep[e.g.,][]{conroy13, courteau14}, and we will discuss the possible impact on our results in Section~\ref{sec:missing_mass} below.

\citet{williams17} found best-fit SFHs using four different model sets to assess the systematic uncertainty due to the choice of stellar evolutionary tracks adopted in the CMD modeling. The four sets of stellar tracks considered are Padova \citep{marigo08, girardi10}, PARSEC \citep[from the same group that produced the Padova models,][]{bressan12}, BaSTI \citep{pietrinferni04, cassisi06, pietrinferni13}, and MIST \citep{choi16}. Each model set adopts a different value of solar metallicity (Padova: $Z_\odot = 0.019$; PARSEC: $Z_\odot = 0.0152$; BaSTI: $Z_\odot = 0.0198$; MIST: $Z_\odot = 0.0142$), so the [M/H] vs. time enforced in the CMD modeling results in different absolute stellar metal mass fractions ($Z$) at a given time across the four fits. We adopt the SFHs derived using the Padova stellar tracks for our fiducial calculations because these results give a total formed stellar mass and mean enrichment history that lie in the middle of the range spanned by the four model sets. 

Figure~\ref{fig:sf_z_history} shows the best-fit histories of star formation (left) and stellar enrichment (right) for these four different sets of stellar models. The fiducial SFHs and enrichment histories using the Padova stellar tracks are shown as solid blue lines, while the differently colored dotted lines show results for the other three model sets: PARSEC (orange), BaSTI (green), and MIST (purple). The left panel plots the total SFR in the PHAT area (summed over all SFH pixels) as a function of the age of the universe, with the corresponding redshift shown on the top axis for reference. Error bars on the Padova SFH show the random uncertainties in the fiducial model fit to the observed CMDs. Random uncertainties for the other model sets are similar and are omitted from the plot for clarity. The model SFR in each age bin is constant, but should be thought of as the average over the duration of each age bin because the SFR is variable over these timescales.

For the fiducial SFHs, the total formed stellar mass within the PHAT area is $M_{\star,\,\mathrm{formed}}^\mathrm{PHAT}=5.01\times10^{10}\,M_\odot$, 78\% of which is formed during the oldest age bin (at $z\gtrsim1$). Finer time resolution is not possible for these old ages because stellar crowding prevents the CMDs from resolving the ancient main-sequence turnoffs. We compute the returned fraction of stellar mass, $R=39.7\%$, from the Padova stellar evolutionary tracks for the total PHAT SFH and enrichment history. Therefore, $1-R = 60.3\%$ of the formed stellar mass is present today, giving $M_\star^\mathrm{PHAT} = 3.02\times10^{10}\,M_\odot$. 

The right panel of Figure~\ref{fig:sf_z_history} shows the enrichment history of the PHAT area for each of the four model sets, with the same color-coding as in the left panel. We plot the mass-weighted mean metal mass fraction ($Z$) of all new stars formed in each time bin against the age of the universe. The four model sets each adopt different solar metal mass fractions, and so a given metallicity relative to solar ([M/H]) corresponds to a different metal mass fraction for each model set. These differing conventions are the main cause of the variation in absolute stellar metal mass fraction across the four different enrichment histories; in particular, the low $Z_\odot$ adopted in the MIST models is the reason why the purple dotted line lies below the rest in the right panel of Figure~\ref{fig:sf_z_history}. For the fiducial (Padova) enrichment history, the mass-weighted mean stellar metallicity over all age bins is $Z=0.013$, or $\mathrm{[M/H]} = -0.16$. 

We use the SFHs and enrichment histories obtained from each of the four different model sets to assess the impact of systematic uncertainty due to the choice of stellar evolutionary tracks on our calculations of stellar metal content (Section~\ref{sec:metals_stars}) and metal retention in M31 (Section~\ref{sec:results}). The random uncertainties on the SFR in each age bin are about 20\%, generally less than the typical variation among SFRs derived for different stellar evolutionary tracks. We use these random uncertainties in to assess the uncertainty in the present metal mass in stars, given the fiducial Padova SFHs (see the left panel of Figure~\ref{fig:metal_retention_total} below). Our goal is to compare the produced and present metal mass, and to be physically consistent, these calculations must use the same SFHs. Therefore, including the random uncertainty in both the produced and current metal mass uncertainty budgets would overestimate the uncertainty in our comparison. The random uncertainty in the SFH is an important contributor to the uncertainty in the current metal mass, but is much smaller than the systematic uncertainties in the metal production model.

\subsubsection{Metal Mass in Stars\label{sec:metals_stars}}

Here, we compute the total metal mass that is present in stars and its spatial distribution. The \citet{williams17} SFHs are broken down into 24 bins of [M/H] at each age, so that the SFR is measured in a grid of age and metallicity. We calculate the stellar mass formed as the SFR in each age and metallicity bin multiplied by the width of the age bin. We then multiply by the metal mass fraction, $Z = Z_\odot \times 10 ^\mathrm{[M/H]}$, to obtain the total metal mass in stars that is formed in each age and metallicity bin.

We use the Padova evolutionary tracks to calculate the fraction of stellar mass that remains locked into stars that are formed at a given age and metallicity, $1-R$. For the old stellar populations ($\gtrsim 5\,\mathrm{Gyr}$ old) that dominate the stellar mass in M31, $\sim60\%$ of the formed mass is present today. A larger fraction of stellar mass remains for younger populations, but these contribute only 17\% of $M_{\star,\,\mathrm{formed}}^\mathrm{PHAT}$. Over all SFH pixels, the fraction of stellar mass remaining, computed for the full SFH, varies only slightly, between 59.1\% and 62.1\%.

\begin{figure*}[!ht]
\minipage{0.5\textwidth}
  \includegraphics[width=\linewidth]{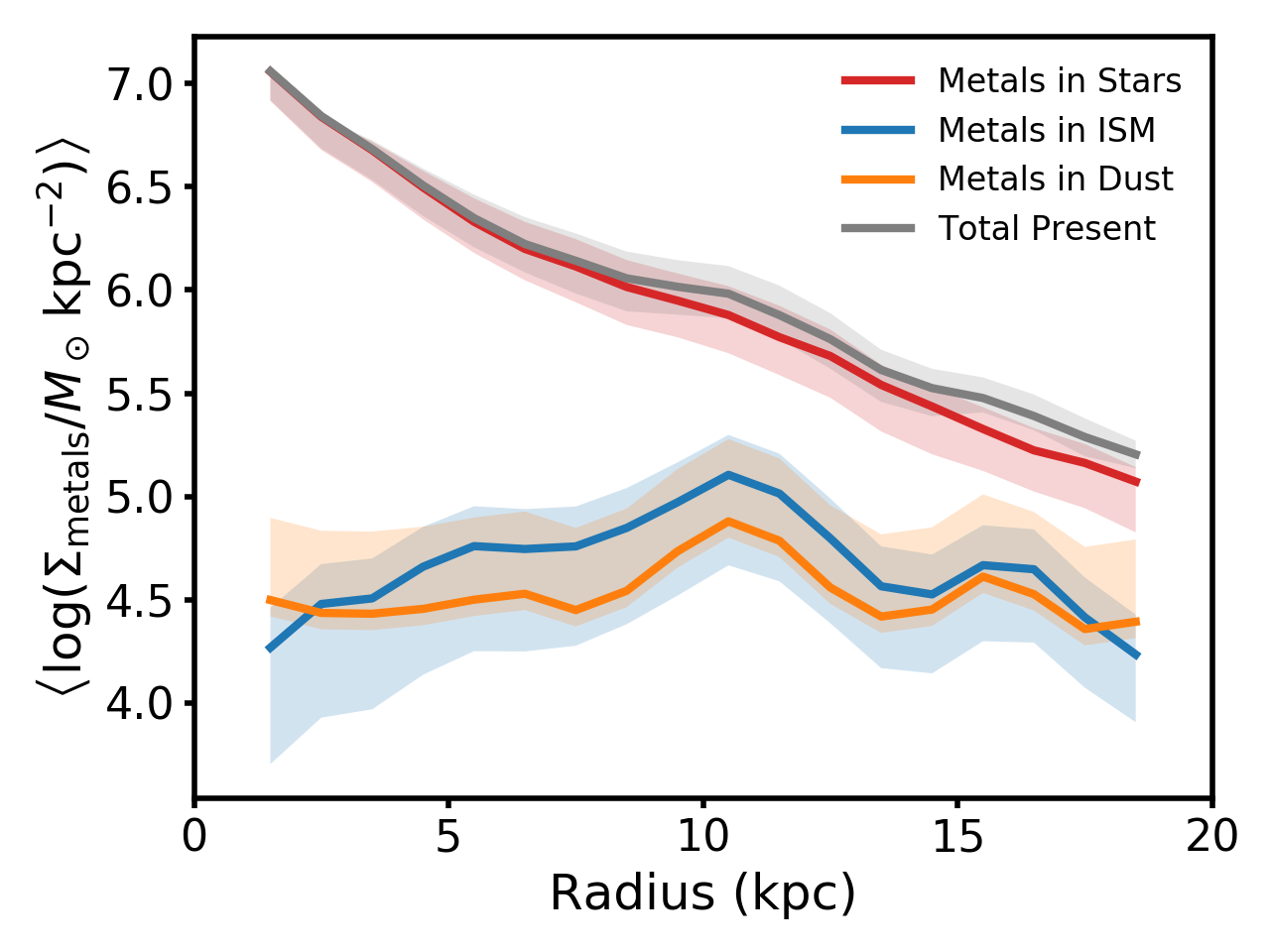}
  \endminipage\hfill
\minipage{0.5\textwidth}
  \includegraphics[width=\linewidth]{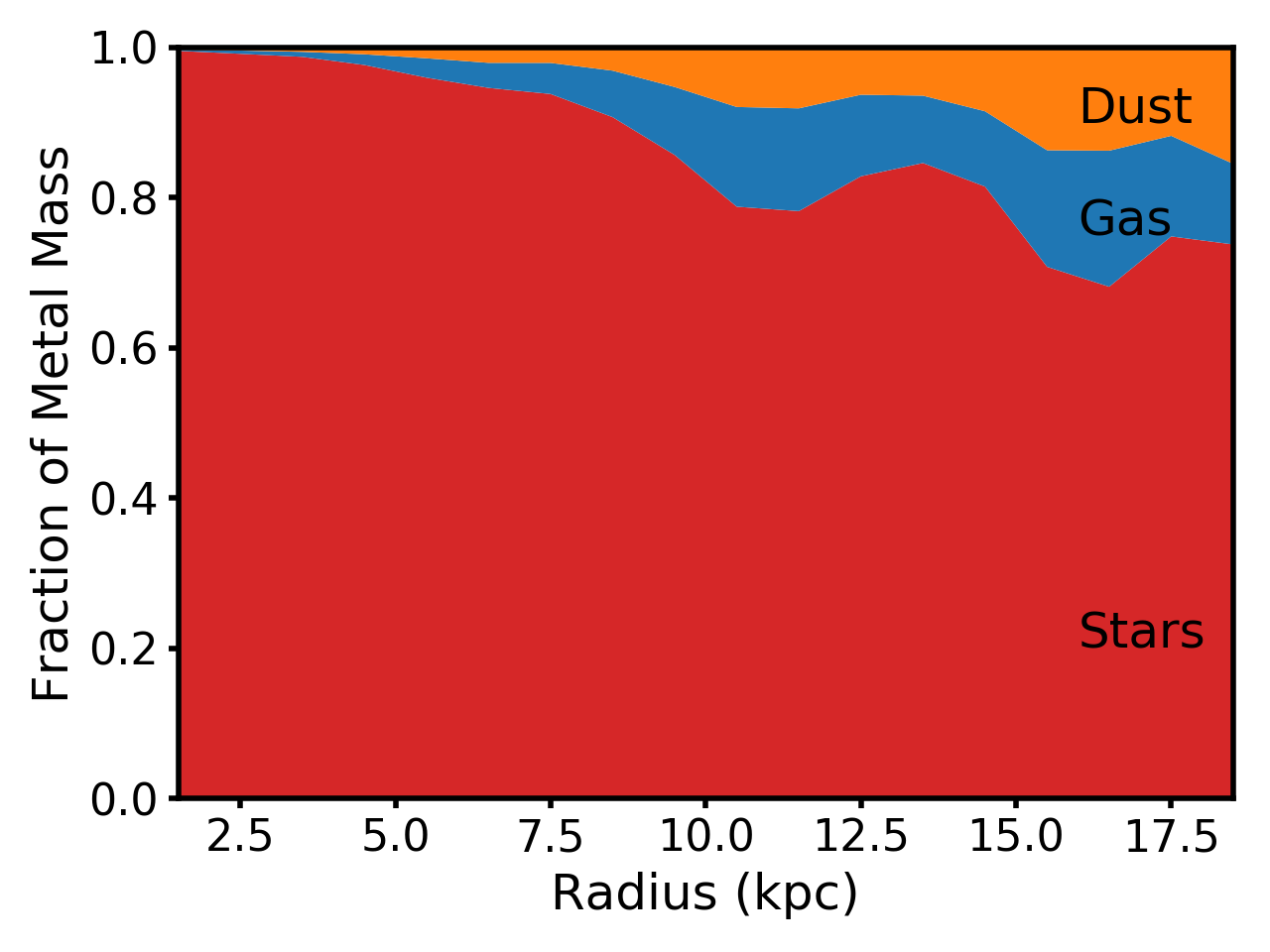}
  \endminipage\hfill
\caption{\textsc{\textbf{The spatial distribution of metals present in M31.}} Left: azimuthally averaged metal mass surface density as a function of radius within the PHAT footprint. The gray solid line shows the total present-day metal mass profile, while the colored lines show the contributions of metals in stars (red), gas (blue), and dust (orange). The shaded regions represent the conservative systematic uncertainty budget for each calculation. Right: fraction of metal mass surface density within each 1 kpc wide annulus contributed by stars, gas, and dust for our fiducial calculation (the solid lines in the left panel). Stars are the dominant metal reservoir at all radii and harbor over 90\% of the metal mass present in the PHAT footprint, although metals in the ISM (gas + dust) contribute up to 30\% of the metal mass in the most gas-rich annuli, tracing the star-forming rings.  
\label{fig:metals_present}}
\end{figure*}

We obtain the metal mass in stars that is currently present in each SFH pixel, $M_\mathrm{metal}^{\star}$, by performing the following summation over all ages and metallicities:

\begin{equation}
\label{eq:stellar_metal_mass}
M_\mathrm{metal}^{\star} =  \sum_Z Z \sum_{\mathrm{age}} \left(1-R(\mathrm{age\,}, Z) \right) M_{\star,\,\mathrm{formed}}(\mathrm{age,\,}Z).
\end{equation}
Integrating $M_\mathrm{metal}^{\star}$ over all SFH pixels, we find $3.9\times 10^8 \,M_\odot$ of metal mass that is currently present in stars. Finally, we calculate the metal mass surface density in stars as $\Sigma_\mathrm{metal}^{\star} = M_\mathrm{metal}^{\star}  / A_\mathrm{pixel}$, where the deprojected pixel area is 0.43 kpc$^{2}$. The radial profile of \sigmametalstar is shown as the solid red line in the left panel of Figure~\ref{fig:metals_present}. 

The dominant source of uncertainty in $\Sigma_\mathrm{metal}^{\star}$ is the choice of stellar evolutionary tracks. We calculate bounding minimum and maximum values of the metal mass content of stars in each SFH pixel using the stellar tracks that give the highest and lowest stellar metal masses: BaSTI and MIST, respectively. Relative to the fiducial calculation using the Padova models, the total metal mass present in stars is 21\% higher for BaSTI and 32\% lower for MIST, corresponding to a range between $2.7$ and $4.8 \times 10^8 \,M_\odot$. The red shaded region in the left panel of Figure~\ref{fig:metals_present} shows the range spanned by these bounding calculations.

Most previous stellar metallicity measurements in M31 used CMD-based techniques, although some spectroscopic metallicities have been measured with limited spatial coverage. \citet{saglia18} used Lick indices measured from spectra of the central $\sim5\,\mathrm{kpc}$ of M31 to constrain stellar population properties. These authors found that the central disk component is of roughly solar metallicity on average, but has a higher metallicity along the bar, and is $\alpha$-enhanced by $\sim0.25\,\mathrm{dex}$ with no dependence on position angle. The \citet{williams17} metallicities are broadly consistent with those found by \citet{saglia18}, although all models used in the CMD modeling were scaled-solar. The \citet{williams17} stellar enrichment histories are also consistent with radial stellar metallicity gradient measured for evolved stellar populations from modeling the RGB in PHAT CMDs \citep{gregersen15}.

\subsection{Gas\label{sec:gas}}

\subsubsection{Maps of the Hydrogen Content\label{sec:gas_maps}}

To calculate $\Sigma_\mathrm{H}$, we consider only the cold ISM gas, the most commonly and easily observed gas phase. Metals certainly reside in the extended hot gaseous halo, but their metallicity is not traced by H \textsc{ii} region abundances. By excluding any hot corona or halo from our metal analysis, we are implicitly leaving that component as a reservoir for any missing metals.

We use the \citet{braun09} map of 21 cm emission that covers the entire disk of M31. The 21 cm observations were taken with the Westerbork Synthesis Radio Telescope (WSRT) and were flux-corrected using single-dish data from the Green Bank Telescope (GBT). We assume that the gas is optically thin throughout and multiply the 21 cm emission map by $1.823 \times 10^{18} \mathrm{cm}^{-2} (\mathrm{K \, km \, s^{-1}})^{-1}$ to calculate the column density $N_\mathrm{H\,I}$. The dominant source of uncertainty in this map is the flux calibration of the single-dish data, and it is at the level of 10-20\%. \citet{braun09} argue that the \hi is optically thick in some regions, meaning that the \hi mass could be higher by up to $\sim30\%$. We therefore include this possibility in our systematic uncertainty budget as a possible upward revision in the total metal mass.

We use the CO$(1-0)$ emission map from \citet{nieten06} as a tracer of the molecular gas phase. The observations were made with the IRAM 30m telescope and cover the central $12\,\mathrm{kpc}$ of M31. The data do not cover the entire PHAT area, but the fractional contribution of $\mathrm{H}_2$ to the total hydrogen mass outside the coverage area is likely to be even lower than the central $12\,\mathrm{kpc}$ value of 12\%. Therefore, our estimate of the hydrogen mass surface density in the outer regions will be at most $\sim10\%$ too low.

We multiply the CO$(1-0)$ emission map by a constant $X_\mathrm{CO} = 2\times10^{20}\,\mathrm{cm^{-2}\, (K \, km \, s^{-1})^{-1}}$ to calculate the molecular hydrogen column density $N_{\mathrm{H}2}$ \citep{bolatto13}. The assumed \xco is the most uncertain ingredient in this calculation, but the level of systematic uncertainty is poorly constrained. We allow for 50\% uncertainty in the range of possible molecular hydrogen mass.

Both the \hi and H$_2$ maps have pixel sizes smaller than the SFH pixels, but we require the average column densities within each SFH pixel for our calculations. We use \textsc{Montage}\footnote{\url{http://montage.ipac.caltech.edu}} (\citealt{berriman03, jacob10}; version 5.0) to reproject the column density maps to the same pixel scale as the \citet{williams17} SFH maps, which is mathematically equivalent to averaging the mass surface density within each SFH pixel. 

Finally, we convert the column density maps into mass surface density maps, converting from cm$^{-2}$ to $M_\odot \, \mathrm{kpc} ^{-2}$ using the known physical size of the SFH pixels and correcting for projection effects. Calculating $\Sigma_\mathrm{H} = \Sigma_\mathrm{H\,\textsc{i}} + \Sigma_\mathrm{H2}$ and summing over all SFH pixels, we find a total of $1.37\times 10^9 \, M_\odot$ of hydrogen within the PHAT footprint. Of this, 87.3\% is in \hi and 12.7\% is in H$_2$; this is consistent with previous findings.

\subsubsection{Metal Mass in the Neutral ISM Gas\label{sec:metals_gas}}

\begin{table*}[!ht]
\begin{centering}
\caption{Parameters of metallicity gradients used to calculate the metal mass that is present in the gas phase.}
\begin{tabular}{lccc}
Calibration             & $12 + \log(\mathrm{O/H})$ $[r=0\,\mathrm{kpc}]$  & Slope $(\mathrm{dex\,kpc}^{-1})$ & $12 + \log(\mathrm{O/H})$ $[r=19\,\mathrm{kpc}]$\\
\hline 
\citet{nagao06} & 8.91 & -0.0195 & 8.54 \\
\citet{pilyugin05} & 8.42 & -0.0054 & 8.32                                    
\end{tabular}
\label{tab:metallicities}
\tablecomments{\citet{nagao06} metallicities have been scaled down by -0.22 dex to avoid double-counting metals in dust.}
\end{centering}
\end{table*}

Gas-phase metallicity ($12 + \log(\mathrm{O/H})$, the abundance of oxygen atoms relative to hydrogen) is measured by modeling the emission lines from H\,\textsc{ii} regions, or by using simple conversions between strong emission line ratios and metallicity that are calibrated using either theoretical models or empirical measurements. In M31, the gas-phase metallicity is consistently found to decrease with radius, but the gradient is quite shallow \citep{zaritsky94, sanders12, zurita12}. The overall normalization of the metallicity is less well constrained, given the systematic differences of up to 0.7 dex between metallicity measurement techniques \citep{kewley08}. 

We calculate the metal content of the ISM gas using the gas-phase metallicity gradients from \citet{sanders12}. They measured gas-phase metallicities of individual H \textsc{ii} regions across the M31 disk using several different strong emission line calibrations and computed the resulting abundance gradients. We have verified that the measured metallicity gradients do not change if they are fit to the metallicities of only the H \textsc{ii} regions that fall within the PHAT footprint. For our fiducial calculations, we adopt the gradient found for metallicities computed using the \citet{nagao06} relation between gas-phase metallicity and [N\,\textsc{ii}]/H$\alpha$, which is available for more H\,\textsc{ii} regions than other line ratios, and which also has the benefit of being insensitive to uncertain dust corrections. Their measured $\log(\mathrm{O/H})$ gradient of $-0.02$ dex kpc$^{-1}$ for this calibrator is consistent with other results in the literature.

The \citet{nagao06} calibration is semiempirical, based on metallicity measurements using the \citet{kewley02} theoretical calibration. These oxygen abundances are scaled up by $0.22$ dex to correct for oxygen depletion onto dust grains \citep{dopita00}, so we scale down the normalization of the \citet{sanders12} [N\,\textsc{ii}]/H$\alpha$ gas-phase metallicity gradient by $-0.22$ dex  to avoid double-counting the metals in dust. We discuss our calculation of metals in dust in Section~\ref{sec:dust} below.

We calculate the number density of oxygen atoms relative to hydrogen in each SFH pixel using the \citet{sanders12} metallicity gradient. The abundance ratio O/H is then converted into a mass ratio by multiplying by the ratio of oxygen to hydrogen atomic masses, and the product of this mass ratio with $\Sigma_\mathrm{H}$ (calculated in Section~\ref{sec:gas_maps} above) gives the oxygen mass surface density.

Finally, we divide by the solar oxygen-to-metal mass ratio from the \citet{anders89} abundance set, $Z(\mathrm{O})/Z = 0.501$, to calculate the total metal mass surface density in the neutral gas phase. In reality, the oxygen-to-metals ratio is not constant in time (1) because the timescales of metal production by different nucleosynthetic sources are different, and (2) because supernova-driven outflows are likely $\alpha$-enhanced. However, there have been no observational results to date showing trends in the oxygen-to-metals ratio in the ISM with galaxy properties. We therefore adopt the constant $Z(\mathrm{O})/Z$ from the solar abundance set that is consistent with the fiducial SFH and emphasize that our reported metal mass in the gas phase is really an oxygen mass scaled to total metal mass by an uncertain constant factor. 

In summary, the metal mass surface density is calculated as 
\begin{equation}
\Sigma_\mathrm{metal}^\mathrm{gas} = \frac{\mathrm{O}}{\mathrm{H}} \times \frac{m_\mathrm{O}}{m_\mathrm{H}} \times \Sigma_\mathrm{H} \times \frac{Z_\odot}{Z(\mathrm{O})_\odot}.
\end{equation} 

The radial profile of \sigmametalgas is shown as the solid blue line in the left panel of Figure~\ref{fig:metals_present}. Integrating over all SFH pixels, we find $2.2\times 10^7 \,M_\odot$ of metal mass currently present in the neutral ISM within the PHAT footprint, which is an order of magnitude lower than the metal mass in stars. Taking the total metal mass in the neutral ISM divided by the total mass in \hi and H$_2$, we find an average metal mass fraction of 0.022, which is higher than any commonly used values of $Z_\odot$, but only 16\% higher than our fiducial $Z_\odot=0.019$. The mass-weighted stellar metal mass fraction is subsolar, 41\% lower than the gas-phase metal mass fraction. This finding is in line with the expectation that the metal content of M31 was lower early in its history when most of its stars were formed.

The choice of metallicity calibration is one of the main uncertainties in the gas-phase metal mass calculation. Because the normalization of the abundance gradient for the \citet{nagao06} [N\,\textsc{ii}]/H$\alpha$ calibration lies at the high-metallicity end of the range, the gas-phase metal content in our fiducial calculation may be biased high. This choice should yield a conservative upper limit on the metal retention fraction and therefore a lower limit on the required metal loss. We calculate a lower bound on the allowed range of metal mass in the neutral ISM using the \citet{pilyugin05} empirically calibrated relation using [O\,\textsc{iii}] and [O\,\textsc{ii}] emission lines, which is known to give metallicities that are systematically low. This metallicity calibrator does not account for metal depletion onto dust grains, so no scaling is required to remove the metals locked into dust from the gas-phase metal budget. Table~\ref{tab:metallicities} provides the parameters of the metallicity gradients used in these calculations.

We calculate a minimum bound on \sigmametalgas assuming the lowest allowed hydrogen mass surface density within the uncertainties described in Section~\ref{sec:gas_maps} and the best-fit radial metallicity gradient for the \citet{pilyugin05} metallicities. Similarly, the maximum bound is calculated using the highest allowed hydrogen content and the metallicity gradient for the \citet{nagao06} metallicities. The resulting range of \sigmametalgas is 65\% lower to 44\% higher than for the fiducial calculation, corresponding to range of metal mass in the gas phase between $7.8\times 10^6$ and $3.2 \times 10^7 \,M_\odot$. The allowed range of \sigmametalgas is shown as the shaded blue region in the left panel of Figure~\ref{fig:metals_present}. For comparison, the scatter in individual $\mathrm{H}\,\textsc{ii}$ region metallicities about the best-fit radial gradient is at the level of $\pm50\%$ for the \citet{nagao06} calibration. 

\subsection{Dust\label{sec:dust}}

\subsubsection{Map of the Dust Content\label{sec:dustmap}}

Finally, we calculate the metal mass in dust grains using a map of the dust extinction, $A_V$, within the PHAT footprint from \citet{dalcanton15}, who modeled the effect of dust on the morphology of the RGB in the NIR CMD. They assumed that the stars in a given $3.3''\times3.3''$ region experience a range extinction and adopted a lognormal probability distribution function (PDF) for $A_V$ (described by a median $\tilde{A_V}$ and spread $\sigma$). The model also includes a reddened fraction, $0 < f_\mathrm{red} < 1$, describing the fraction of stars in each region that lies behind the dust layer, while the remainder of the stars are in front of the dust and therefore experience no attenuation. This approximation is appropriate for evolved stellar populations whose scale height is large relative to the thin dust layer. 

\citet{dalcanton15} produce a high-resolution map of the best-fit median $\tilde{A_V}$, from which we calculate the mean $A_V$ in each SFH pixel. The total extinction PDF within the PHAT footprint peaks at $A_V = 0.96$, with a long tail to higher values. The central 95\% of $\tilde{A_V}$ spans the range $0.05-2.1$. Typical uncertainties on $\tilde{A_V}$ are at the 20\% level, but reach up to 50\% in low-dust regions where the dust model parameters are poorly constrained. These more uncertain low-dust regions contribute only a small amount to the total dust mass budget.

\subsubsection{Metal Mass in Dust\label{sec:metals_dust}}

Dust extinction scales linearly with the dust mass surface density. We use the scaling given by \citet{draine07} to calculate \sigmadust from the average $A_V$ measured within each SFH pixel. However, this dust model has been found to predict extinctions in M31 that are $\sim2.5\times$ higher than the $A_V$ maps found by other authors using different measurement techniques \citep{dalcanton15, planck16-dust}. It remains unclear whether this systematic offset is driven by uncertainties in the conversion from IR luminosity to \sigmadust, or from \sigmadust to $A_V$. If the former, then our calculation of \sigmadust from the $A_V$ map would be unaffected, but if the latter, then we should scale down our \sigmadust obtained from the \citet{draine07} model.

We adopt the constant renormalization recommended by \citet{dalcanton15}, scaling down \sigmadust from the \citet{draine07} model by a factor of 2.5 in our fiducial calculation. Including this renormalization, we calculate $\Sigma_\mathrm{dust}$ as 
\begin{equation}
 \Sigma_\mathrm{dust} = 5.41 \times 10^{4} M_\odot \, \mathrm{kpc}^{-2} \left(A_V / \mathrm{mag}\right).
\end{equation}
This choice gives an integrated gas-to-dust ratio of 99.8, consistent with previous measurements for the northern disk of M31 \citep{leroy11}. We assume that dust is entirely composed of metals, such that $\Sigma_\mathrm{metal}^\mathrm{dust} = \Sigma_\mathrm{dust}$. The radial profile of \sigmametaldust is shown as the solid orange line in the left panel of Figure~\ref{fig:metals_present}. We find a total of $1.4 \times 10^7 \, M_\odot$ of metal mass in dust within the PHAT footprint.

We calculate a lower bound on \sigmametaldust by allowing the $A_V$ measurements to decrease by 20\%, the minimum allowed within the measurement uncertainties. For the upper bound, we assume that  \sigmametaldust obtained from the \citet{draine07} model is correct without the factor of 2.5 reduction. These choices give a range of possible \sigmametaldust  20\% lower to 150\% higher than found for the fiducial calculation, corresponding to a total present-day metal mass in dust between $1.1$ and $3.6 \times 10^7 \, M_\odot$. The allowed range of \sigmametaldust is shown as the shaded orange region in the left panel of Figure~\ref{fig:metals_present}. Although this is a large range, the majority of metal mass that is present in the M31 disk is in stars, so the dust mass uncertainty does not dominate the systematic uncertainty budget.

\subsection{Total Current Metal Mass and Its Distribution\label{sec:metals_present}}

As referenced above, we present the radial profiles of the metal mass surface density and the contribution of metals in stars, gas, and dust in the left panel of Figure~\ref{fig:metals_present}. All surface densities are averaged in 1 kpc wide annuli. The solid gray line shows the radial profile of the total metal mass surface density, while the colored lines show the radial profiles for each metal reservoir: \sigmametalstar (red), \sigmametalgas (blue), and \sigmametaldust (orange). The shaded regions show the minimum and maximum bounds allowed within the systematic uncertainties (described in Sections~\ref{sec:gas}--\ref{sec:dust} above).

The right panel of Figure~\ref{fig:metals_present} shows the fraction of metal mass surface density residing in each metal reservoir as a function of radius, with the same color-coding as in the left panel. Only the results for the fiducial calculation are shown. \sigmametalstar contributes most to the total metal mass surface density at all radii, but especially in the inner $\sim8\,\mathrm{kpc}$. The gas surface density is very low in this central region and does not harbor many metals, so over 90\% of metal mass is in the stellar component. In the outer regions, however, the gas and dust contribute up to 35\% of the metal mass surface density in the most gas-rich annuli.

We find that a total of $4.3\times10^8\,M_\odot$ of metal mass is present in the PHAT region of M31 for our fiducial calculations. When we account for random errors in the fiducial Padova SFH and systematic uncertainties in the gas-phase metallicity, hydrogen mass, and dust content, the allowed range of present metal mass in the PHAT area is $3.3-5.4\times10^8\,M_\odot$. Of the fiducial present metal mass, 91.7\% is in stars, 5.1\% is in the neutral ISM, and 3.2\% is in dust. The stellar mass surface density is highest in the central regions, where the metal mass in stars dominates over the other components. So even though the neutral gas and dust are important metal reservoirs in the outer annuli, the metals in stars dominate the present-day metal census in M31. 

\citet{peeples14} have performed a similar accounting of metal mass in stars, gas, and dust across the local galaxy population. Their analysis employs scaling relations between metallicity, gas mass, and dust content and galaxy stellar mass to estimate the total metal mass present in the stars, gas, and dust as a function of stellar mass. This is a powerful approach to measuring metal retention in a statistical sense, but is limited by the intrinsic scatter in the various scaling relations employed. We compare our measured present and produced metal mass in M31 to estimates from the statistical analysis of \citet{peeples14}.

For a galaxy of M31's stellar mass ($10^{11}\,M_\odot$), \citet{peeples14} find $1.7\times10^9 \,M_\odot$ of metals present. We calculate the present metal mass in the M31 disk by scaling up the metal mass currently in the PHAT area, assuming azimuthal symmetry and extrapolating the metal production surface density between $1$ and $2\,\mathrm{kpc}$ inward to account for the contribution of the central $1\,\mathrm{kpc}$. We find $1.1\times10^9\,M_\odot$ of metals in the disk of M31 ($r<19\,\mathrm{kpc}$), with a possible range between $8.7\times10^8$ and $1.4\times10^9\,M_\odot$ within our uncertainty budget. The entire possible range of present metal mass is below the \citet{peeples14} value, 35\% lower for our fiducial calculation. The fraction of metal mass found in stars, gas, and dust by \citet{peeples14} is 75\%, 15\%, and 10\%, respectively. These differences suggest that M31 would be an outlier from the average metal census of \citet{peeples14}, particularly in terms of the relative importance of the various metal reservoirs.


\section{The Spatially and Temporally Resolved History of Metal Production\label{sec:metal_history_all}}

In this section, we calculate the history of metal production in M31 and the total metal mass produced. We introduce our model of metal production following a burst of star formation and discuss the main sources of systematic uncertainty. We combine the metal production model with the spatially resolved Padova SFHs (described in Section~\ref{sec:sfhs}) to calculate the history of metal production by each nucleosynthetic source. We assess the effect of using SFHs derived using different stellar evolutionary tracks on the total metal production. Finally, we assume azimuthal symmetry to scale up the metal production within the PHAT footprint to determine the total metal mass that was produced in the M31 disk (at $r < 19\,\mathrm{kpc}$).

\subsection{Model of Metal Production\label{sec:metal_model}}

Here, we describe our fiducial model of metal production by Type II supernovae (SNe), AGB stars, and Type Ia SNe in turn. We discuss how we bound the range of possible metal yields due to the main sources of systematic uncertainty in each calculation. These three nucleosynthetic sources account for most of the newly formed metal mass because they dominate the production of the most abundant elements. More exotic and rare processes (e.g., neutron star mergers) also produce new metals, but these events typically dominate the production of certain rare elements. Because they do not contribute much to the total metal mass, neglecting rarer metal production sources does not affect our results.

For each nucleosynthetic source, we construct a model of the metal production rate as a function of time following a star formation event. For Type II SNe and AGB stars, we assume that new metals are returned to the ISM at the end of the star's lifetime. We assume that the lifetime of a $1\,M_\odot$ star is 10 Gyr and adopt a power-law relation between stellar mass and age \citep{prialnik09}, 
\begin{equation}
t / 10 \mathrm{\,Gyr} = (M_\star / M_\odot) ^ {-2.6}
\end{equation}

Throughout, we adopt a \citet{kroupa01} IMF for consistency with the assumptions made in deriving the SFHs (see Section~\ref{sec:sfhs}). The choice of IMF strongly affects the predicted metal yields because the IMF dictates the number of high-mass stars that are available to produce Type II SNe, which dominate the overall metal production, per solar mass of new stars that are formed. It is known that varying the choice of IMF causes the metal yield to vary by up to a factor of three \citep{vincenzo16}, such that shallower slopes at the low-mass end \citep[e.g., ][]{kroupa01, chabrier03} result in more metal production per unit mass of formed stars. However, if an IMF that produced more low-mass stars were adopted in the CMD modeling \citep[e.g.,][]{salpeter55}, then the recovered SFHs would change such that the overall stellar mass that was formed increased. The decreased metal production per mass of stars formed for that steeper low-mass IMF would roughly counteract the increase in overall stellar mass formed. Because the CMD-based SFHs that we have adopted were derived assuming a \citet{kroupa01} IMF, we adopt the same IMF throughout this work to maintain consistency, and do not include the uncertainty in the IMF in our systematic uncertainty budget.

In this paper, we define the metal yield, $y$, as the ratio of newly produced metal mass to the formed stellar mass: $y = M_Z^\mathrm{produced} / M_\star^\mathrm{formed}$.  Our quoted ``metal yields'' are not directly comparable to some other papers in the literature, which often use ``net yields'' or ``yields per stellar generation" \citep[e.g.,][]{vincenzo16}. These definitions differ by a factor of $1/(1-R)$, where $R$ is the fraction of stellar mass returned to the ISM, to scale the total metal yield to present-day stellar mass: $y_\mathrm{net} = y / (1-R)$. 

\begin{figure*}[!ht]
\minipage{0.5\textwidth}
  \includegraphics[width=\linewidth]{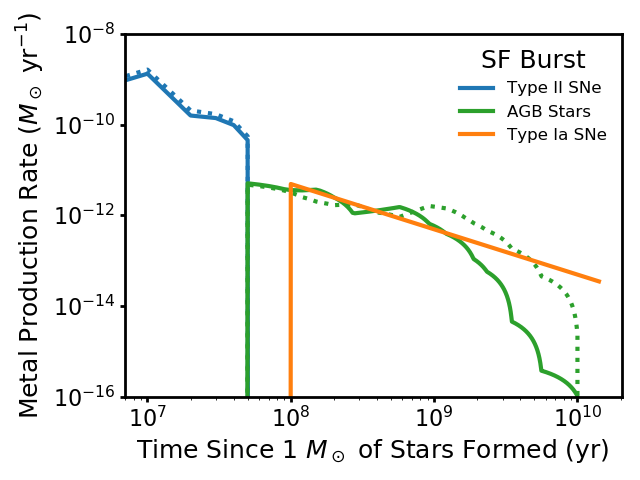}
\endminipage\hfill
\minipage{0.5\textwidth}
  \includegraphics[width=\linewidth]{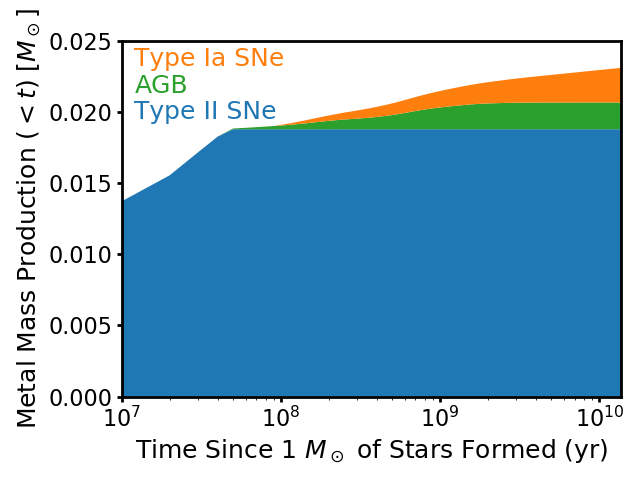}
\endminipage\hfill
\caption{\textsc{\textbf{Fiducial model of metal production following a burst of star formation.}} Left: the metal mass production rate, including all metal species, from Type II SNe (blue), Type Ia SNe (orange), and AGB stars (green) as a function of time following an instantaneous burst of star formation, normalized to $1\,M_\odot$ of stars formed. The solid lines show our fiducial model (adopting stellar metallicity $Z=0.008$, similar to the mass-weighted mean stellar metallicity in M31), while the dotted lines illustrate how metal production by Type II SNe and AGB stars changes for the lowest stellar metallicity models ($Z=0.0001$). The shaded regions in Figure~\ref{fig:metal_history} below illustrate the range of possible metal production histories due to systematic uncertainty in this metal production model. Right: the cumulative metal mass formed up to a given time following the burst by each nucleosynthetic source for the fiducial model (the solid lines in the left panel). Type II SNe dominate the overall metal production, but all new metal mass that is produced more than 50 Myr after the burst is due to AGB stars and Type Ia SNe. 
\label{fig:metal_model}}
\end{figure*}

\subsubsection{Type II Supernovae\label{sec:typeii}}

High-mass stars ($\gtrsim 8\,M_\odot$) produce most new metal mass as they end their lives, expelling metals in winds and during their explosive deaths as Type II SNe. These explosive events are also important sources of energetic feedback to the ISM, driving turbulence and even removing gas from regions where vigorous star formation is ongoing. Type II SNe dominate the production of  $\alpha$-elements (e.g., O and Si) which are often used to trace the physical state of gas within galaxies.

We adopt metal yields from the \citet{nomoto13} stellar models, which are available for several discrete initial stellar masses in the range of $13-40\,M_\odot$ and for four metallicities ranging from $Z=0.001-0.02$. Metallicity weakly affects the metal production by Type II SNe, in the sense that the lowest metallicity models produce $\sim20\%$ more metal mass than the highest metallicity models. We construct metal production models for each metallicity to properly account for variation in metal production due to the enrichment history of M31 in Section~\ref{sec:metal_history} below, but for simplicity, we present here a fiducial model with $Z=0.008$, which is closest to the mass-weighted mean metallicity of stars in M31.

We assume that stars between $8$ and $40\,M_\odot$ explode as Type II SNe and return metals to the ISM, so we extend the $13\,M_\odot$ yield down to $8\,M_\odot$. Although stars more massive than $40\,M_\odot$ likely do explode as SNe, these most massive stars probably form black hole remnants. Our fiducial model assumes that any newly produced metals fall back onto these remnants and never return to the ISM; this is a standard assumption in chemical evolution modeling \citep[e.g.,][]{poulhazan18}. 

The fiducial metal production rate by Type II SNe as a function of time following a burst of star formation is shown as the solid blue line in the left panel of Figure~\ref{fig:metal_model}. The metal production rate is normalized to $1\,M_\odot$ of formed stellar mass. For reference, the lowest metallicity Type II SNe metal production model is shown as the dotted blue line. The total metal mass produced up to a given time by Type II SNe is shown as the solid blue shading in the right panel of Figure~\ref{fig:metal_model}. All metal production by Type II SNe occurs within 50 Myr of the burst. The metal yield for Type II SNe is 0.01878 for this model, or 81.3\% of the total metal yield over a Hubble time.

The main uncertainties in the metal production by Type II SNe are (1) the highest stellar mass star that returns metals to the ISM \citep{vincenzo16}; and (2) the variation among theoretical yields, and in particular, whether the effects of winds and rotation are taken into account \citep{romano10}. The \citet{nomoto13} metal yields we adopted do not include stellar rotation or pre-supernova metal production, and so may underestimate the total metal production by high-mass stars.

To calculate a minimum bound on the Type II SN metal production, we assume that all stars above $25 \, M_\odot$ collapse to black holes and do not return metals to the ISM (following \citealt{emerick18}), and use the highest metallicity model. For the maximum bound, we allow stars up to $100\,M_\odot$ to return metals to the ISM (following \citealt{oppenheimer08, peeples14}), use the lowest metallicity model, and increase the metal yields by 30\% to account for the typical impact of stellar rotation and pre-SN metal production \citep{romano10}. Our resulting allowable range of Type II SNe metal yields is between 48\% lower and 132\% higher than the fiducial model. These minimum and maximum Type II SNe metal production models bound the shaded blue region in the left panel of Figure~\ref{fig:metal_history}.

\subsubsection{AGB Stars\label{sec:agb}}

Low- and intermediate-mass stars become AGB stars and undergo substantial mass loss as they end their lives. These stars return new metals to the ISM over much longer timescales following a burst of star formation than Type II SNe, as even $\sim1\,M_\odot$  stars (with 10 Gyr lifetimes) go through an AGB phase. The total metal mass produced by AGB stars is small compared to that produced by Type II SNe, but AGB stars produce a large fraction of a few common elements, including C and N. Total metal production (and the relative production of different elements) during the AGB phase depends strongly on the initial metallicity of the star. 

We use the metal yields from the AGB star models of \citet{karakas10}, spanning initial stellar masses $1-6 \, M_\odot$ and metallicities $Z=0.0001-0.02$. AGB metal production is more sensitive to the stellar metallicity than for Type II SNe. Again, we construct metal production models for each available metallicity and properly account for the enrichment history when calculating the metal production history of M31 in Section~\ref{sec:metal_history} below, but discuss the $Z=0.008$ model here. We assume that stars with initial masses of $1-8 \,M_\odot$ become AGB stars and return metals to the ISM, but the \citet{karakas10} models only go up to $6 \,M_\odot$. We therefore extend the yield of the $6 \,M_\odot$ model up to $8 \,M_\odot$. 

We show the metal production rate by AGB stars with initial metallicity $Z=0.008$ as the green solid line in the left panel of Figure~\ref{fig:metal_model}. For comparison, the metal production by metal-poor AGB stars ($Z=0.0001$) is shown as as the dotted green line. Metal production by AGB stars begins when Type II SNe metal production ends, at 50 Myr after a star formation episode, and continues up to 10 Gyr. Lower mass AGB stars produce less new metal mass per star, but because they are more numerous than higher mass stars, they contribute appreciably to new metal production long after the star formation episode.

The metal yield from AGB stars up to a given time is shown as the green shaded region in the right panel of Figure~\ref{fig:metal_model}. This plot demonstrates that the contribution of AGB stars to the total metal mass produced is small compared to that of Type II SNe, and most AGB metal mass is returned in the first few gigayears following the star formation episode (see the flattening of the green shaded region after $\sim10^9\,\mathrm{yr}$). The metal yield for the fiducial ($Z=0.008$) AGB model is 0.00187, or 8.1\% of the total metal yield. 

Theoretical AGB metal yields are highly sensitive to the treatment of convection
and mass loss in the modeling \citep[e.g.,][]{karakas14}. The metal yields calculated using different modeling techniques vary \citep[e.g.,][]{marigo01, gavilan05}, but a comprehensive comparison of AGB metal yields is beyond the scope of this paper. Across all AGB models, the total metal production is highly sensitive to the initial metallicity of the star at a level that is comparable to the variation across models that employ different treatments of uncertain processes. We therefore assess the systematic uncertainty in AGB metal production by changing the initial metallicity of the stars in M31. We calculate very conservative minimum and maximum bounds by fixing the initial stellar metallicity to the maximum ($Z=0.02$) and minimum ($Z=0.0001$) values, respectively, for which models are available from \citet{karakas10}. This results in allowed AGB metal yields ranging from 60\% lower to 84\% higher than the $Z=0.008$ model, which is most similar to the mass-weighted mean stellar metallicity in M31. These minimum and maximum AGB metal production models are shown as the shaded green region in the left panel of Figure~\ref{fig:metal_history}. 

\subsubsection{Type Ia Supernovae\label{sec:typeia}}

The final nucleosynthetic source in our model is Type Ia SNe, which dominate the production of the iron-peak elements (e.g., Fe, Ni). It is thought that the progenitors of Type Ia SNe are white dwarfs (WDs), formed from stars that were initially $\sim3-8 \, M_\odot$. However, the exact mechanism that produces the explosion is still under debate; theoretical models and observations have yet to converge on a consistent picture. The Type Ia SN rate as a function of time following a burst of star formation is called the delay time distribution (DTD). There is much ongoing observational effort to constrain the form of the DTD and the time interval following a burst during which Type Ia SNe explode.

We take an empirical approach and adopt a DTD based on a compilation of measurements from \citet{maoz14}. The power-law slope ($t^{-1}$) and normalization ($4\times10^{-13}\,\,\mathrm{SN\,yr}^{-1}M_\odot^{-1}$) from \citet{maoz12a} lie roughly in the middle of the range of DTDs found by various authors; this slope is in line with theoretical expectations for the ``double degenerate'' scenario where two WDs merge to produce the SN explosion. We assume that the first Type Ia SN occurs 100 Myr after a star formation episode \citep{schawinski09, anderson-j15}, roughly the lifetime of a $5.5\,M_\odot$ star. To calculate the metal production rate, we adopt the metal yield per Type Ia SN from the W7 deflagration models of \citet{tsujimoto95}. Because the nature of Type Ia SNe is still poorly understood, we do not attempt to model the effect of progenitor metallicity on new metal production.

The fiducial Type Ia SNe metal production rate as a function of time following a starburst is shown as the orange line in the left panel of Figure~\ref{fig:metal_model}. Again, the total metal mass produced by Type Ia SNe, normalized to $1\, M_\odot$ of stars formed, up to a given time is shown as the orange shaded region in the right panel. The total metal yield over a Hubble time is 0.00245, or 10.6\% of the total metal yield in the fiducial model. Our model does not include a break in the power-law DTD at long delay times following the star formation episode \citep[this possibility was disfavored by][]{heringer17}, and so even the earliest star formation episodes continue to produce Type Ia SNe, and therefore new metal mass, at a low rate up to the present day.

To bound our minimum and maximum allowed Type Ia SNe metal production, we vary the parameters describing the DTD: (1) the time after the star formation event when the first Type Ia SN explodes (45--100 Myr, the lifetimes of 8 and 5.5 $M_\odot$ stars); (2) the power-law slope ($t^{-1}$ for the minimum and $t^{-1.5}$ for the maximum model, following \citealt{heringer17}); and (3) the normalization ($1/4-2$ times the fiducial value, to span the range of observational constraints presented in \citealt{maoz14}). The resulting range of allowed metal yields due to Type Ia SNe is between 75\% lower and 255\% higher than the fiducial calculation. These extreme Type Ia SNe metal production models bound the shaded orange region in the left panel of Figure~\ref{fig:metal_history}.

\subsection{The History of Metal Production in M31\label{sec:metal_history}}

\begin{figure*}[!ht]
\minipage{0.5\textwidth}
  \includegraphics[width=\linewidth]{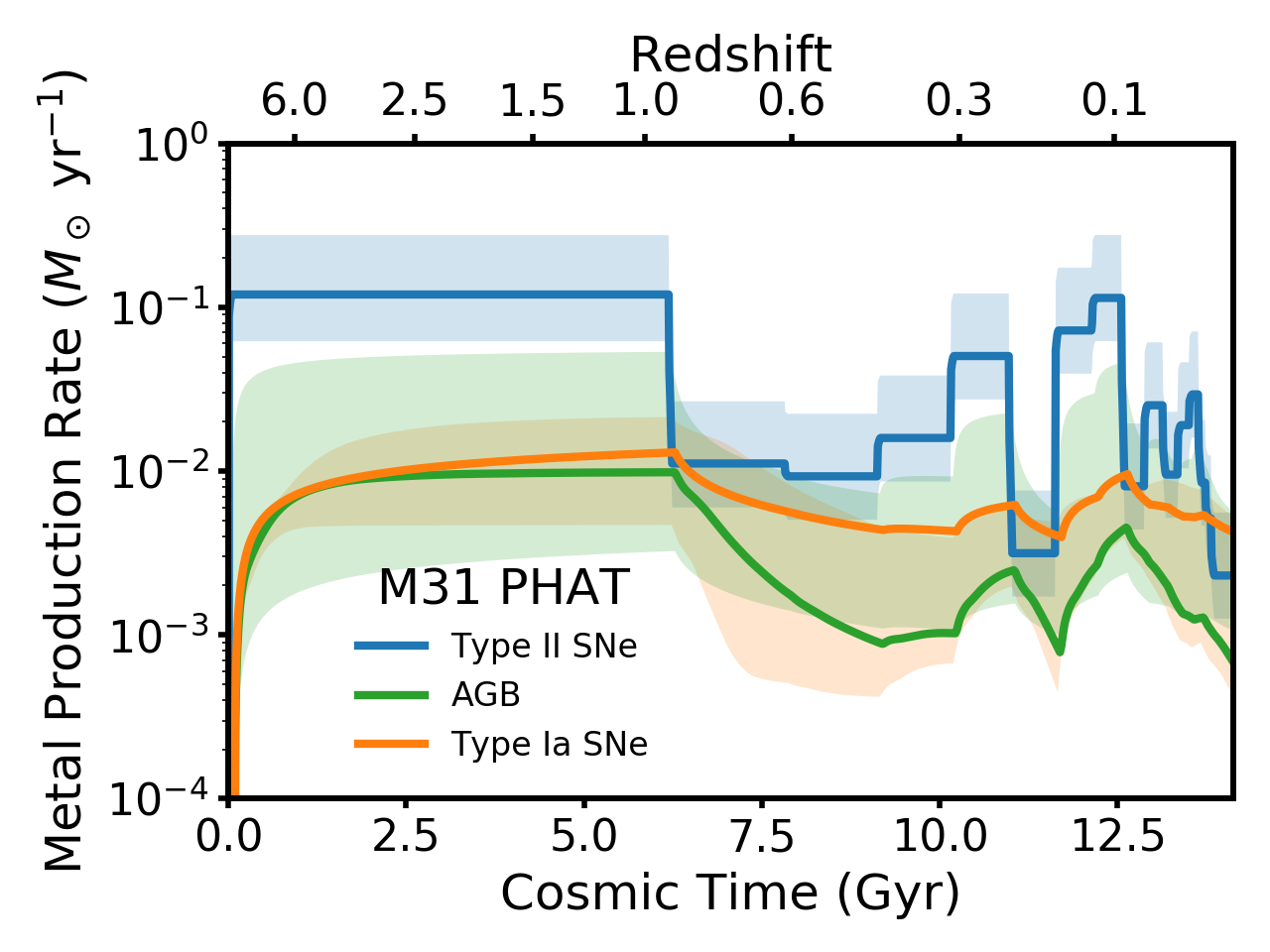}
\endminipage\hfill
\minipage{0.5\textwidth}
  \includegraphics[width=\linewidth]{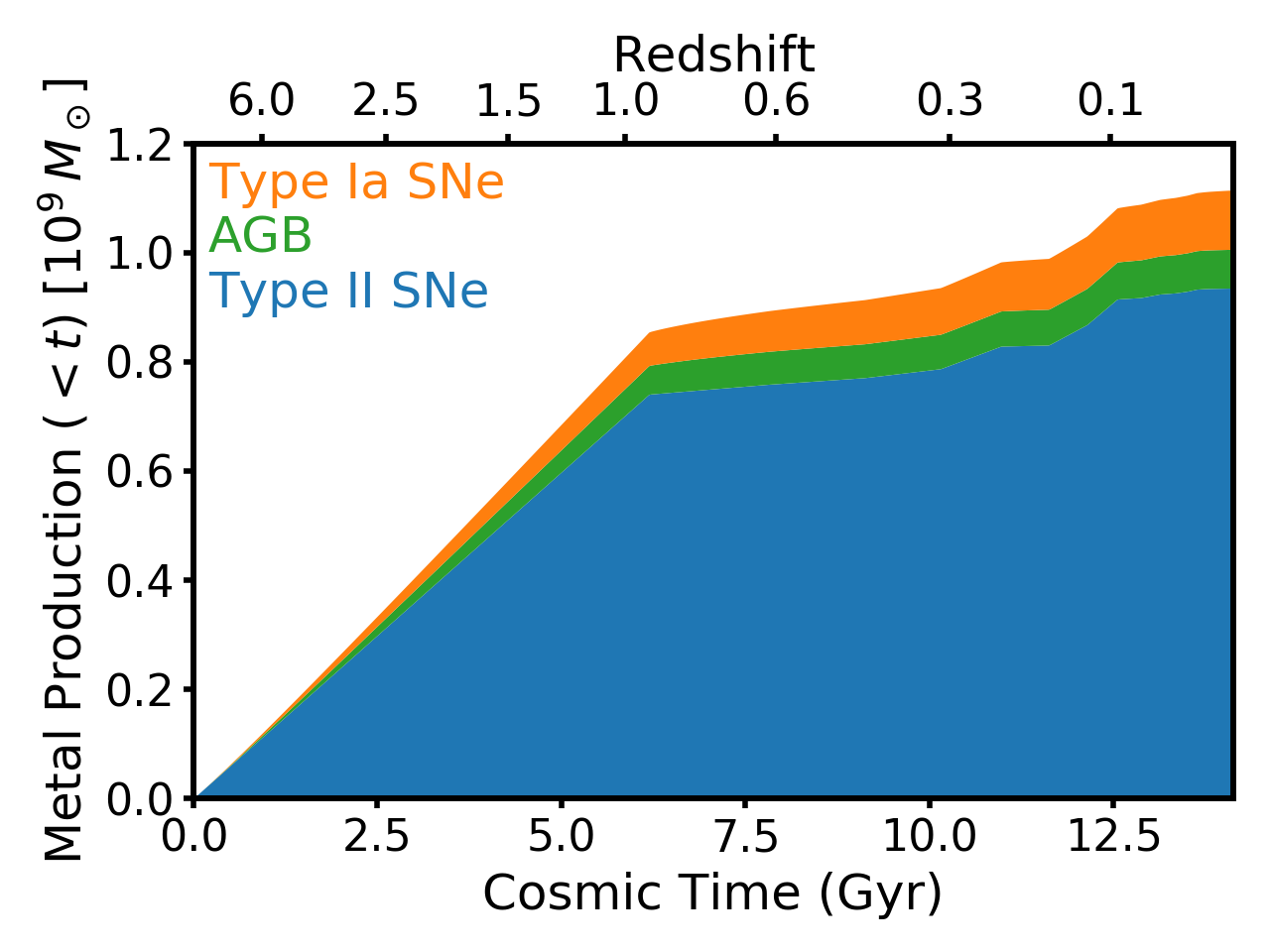}
\endminipage\hfill
\caption{\textsc{\textbf{The history of metal production in M31.}} Left:  the total metal production rate within the PHAT footprint due to Type II SNe (blue), Type Ia SNe (orange), and AGB stars (green) as a function of the age of the universe. Solid lines show the metal production history for the fiducial model, adopting the Padova SFHs and tracking the contributions of stars that formed at different metallicities. The shaded regions illustrate the conservative range of systematic uncertainties due to the choice of parameters in the metal production model (discussed in Sections~\ref{sec:typeii}--\ref{sec:typeia}). Right: the cumulative metal mass produced within the PHAT footprint up to a given age of the universe for the fiducial model (the solid lines in the left panel). Most metal production occurred early, with 77\% of metal mass formed in the oldest age bin ($z\gtrsim1$). Type II SNe dominate the metal production, but the fractional contribution of the delayed metal production by AGB stars and Type Ia SNe increases from $z\sim1$ to the present day. 
\label{fig:metal_history}}
\end{figure*}

Here, we combine the metal production model (Section~\ref{sec:metal_model}, Figure~\ref{fig:metal_model}) above with the spatially resolved SFHs from PHAT (Section~\ref{sec:sfhs}, Figure~\ref{fig:sf_z_history}) to calculate the metal production history in each SFH pixel. Stars formed in each age bin span a range of metallicity, which affects the metal production by Type II SNe and AGB stars. To properly track the contribution to metal production by stars of different metallicities, we use our knowledge of the distribution of stellar metallicity in each time bin in the SFHs (described in Section~\ref{sec:sfhs} above). We divide the SFHs into four bins of metallicity, where each bin is centered on one of the four values of $Z$ sampled by the \citet{karakas10} and \citet{nomoto13} models. We then convolve the SFH in each metallicity bin with the appropriate metal production model for that metallicity. This division ensures that we capture the changing metal production by AGB stars and Type II SNe as the galaxy becomes more metal rich. 

The left panel of Figure~\ref{fig:metal_history} presents the resulting total metal production rate as a function of time. The colored solid lines show the fiducial metal production history by each nucleosynthetic source, where the color-coding is the same as in Figure~\ref{fig:metal_model}. The shaded regions show the range of metal production histories allowed by considering extreme choices that one could make in the metal production model (described in Sections~\ref{sec:typeii}--\ref{sec:typeia} above).

The cumulative metal mass produced by each nucleosynthetic source up to a given time is shown in the right panel of Figure~\ref{fig:metal_history}, with the same color-coding as in Figure~\ref{fig:metal_model}. Metal production is dominated by Type II SNe, and most metal mass is produced during the oldest age bin, at $z\gtrsim1$. Up to the present day, $1.1\times10^{9}\,M_\odot$ of metal mass has been produced within the PHAT footprint, with 83.9\% of that mass produced by Type II SNe, 9.7\% by Type Ia SNe, and 6.4\% by AGB stars. These fractional contributions are slightly different than those quoted for the $Z=0.008$ model in Section~\ref{sec:metal_model} above for two reasons: (1) low-metallicity stars contribute more to AGB metal production, and (2) not all Type Ia SNe that are due to intermediate-age and recent star formation have exploded yet. 

For the Padova SFHs, between $5.5\times10^{8}\,M_\odot$ and $2.8\times10^{9}\,M_\odot$ of metals may have been produced (between 50\% lower and 148\% higher than in the fiducial model) inside the PHAT footprint within the allowed range of systematic uncertainties in the metal production model. The total metal mass that is produced does change for SFHs derived using different stellar evolutionary tracks (PARSEC, BaSTI, or MIST; see Section~\ref{sec:sfhs}), because each model set results in a different enrichment history and total stellar mass formed. For the fiducial metal production model, but using SFHs derived using different stellar evolutionary tracks, we find only modest differences from the fiducial Padova case: the total metal production is just 13\% higher for either BaSTI or MIST, and it is essentially unchanged for PARSEC.

We compare the metal production in the M31 disk for our fiducial model to the expected metal mass produced by a galaxy of M31's stellar mass calculated by \citet{peeples14}. Assuming azimuthal symmetry and extrapolating the metal production surface density between $1$ and $2\,\mathrm{kpc}$ inward to account for the contribution of the central $1\,\mathrm{kpc}$, we scale up the metal mass produced within the PHAT footprint and find that $2.9\times10^9 \,M_\odot$ of metal mass was produced in the entire M31 disk ($r < 19 \,\mathrm{kpc}$). Within our conservative systematic uncertainty budget, between $1.4\times10^9$ and $7.2\times10^9\,M_\odot$ of metals may have been produced. Our fiducial total metal mass produced is lower than half the value from \citet{peeples14}, who found $6.3\times10^9\, M_\odot$ of metal mass produced by a galaxy with $M_\star \sim 10^{11}\,M_\odot$. This discrepancy can be explained by different choices in the metal production models used in their work relative to ours, and is within the level of systematic uncertainty that we report in our metal production model. 

Figure~\ref{fig:metal_history} demonstrates that the relative contribution of different nucleosynthetic sources to new metal production changes over time. In particular, when the SFR drops from one age bin to the next (e.g., at the end of the oldest age bin), the Type II SNe contribute a smaller fraction of the new metals because the delayed Type Ia SNe and AGB star metal production rates are still elevated due to the higher SFR before the quenching event. Following a sudden drop in the SFR, this might change the abundance ratios in the ISM (elevating C, N, and Fe relative to $\alpha$ elements) from would be expected for a constant SFR, and these changes could persist over gigayear timescales if the SFR remains low. However, the magnitude of the change in abundance ratios depends on the amount of previously formed metal mass that is retained in the ISM. Most metals are produced at early times by Type II SNe, so if all of those metals are retained, the ISM abundance ratios will never deviate much (in a spatially averaged sense) from those of Type II SNe ejecta. Observational evidence of spatially coherent enhancements of AGB and Type Ia SNe nucleosynthetic products in ISM and/or stellar populations may therefore be leveraged as evidence of metal loss events.


\section{Results: Metals Missing from M31\label{sec:results}}

Here, we compare the total metal mass produced within the PHAT footprint to the current metals present in the same area. We then calculate the total metal mass missing from the M31 disk (at $r < 19\,\mathrm{kpc}$) under the assumption of azimuthal symmetry and present both the total and spatially resolved metal retention fraction, $f_\mathrm{retained}$. Throughout, we discuss the robustness of these results to systematic uncertainties and to possible confounding effects of stellar radial migration and hierarchical assembly. 

\subsection{Total Missing Metal Mass\label{sec:missing_mass}}

\begin{figure*}[!ht]
\minipage{0.55\textwidth}
  \includegraphics[width=\linewidth]{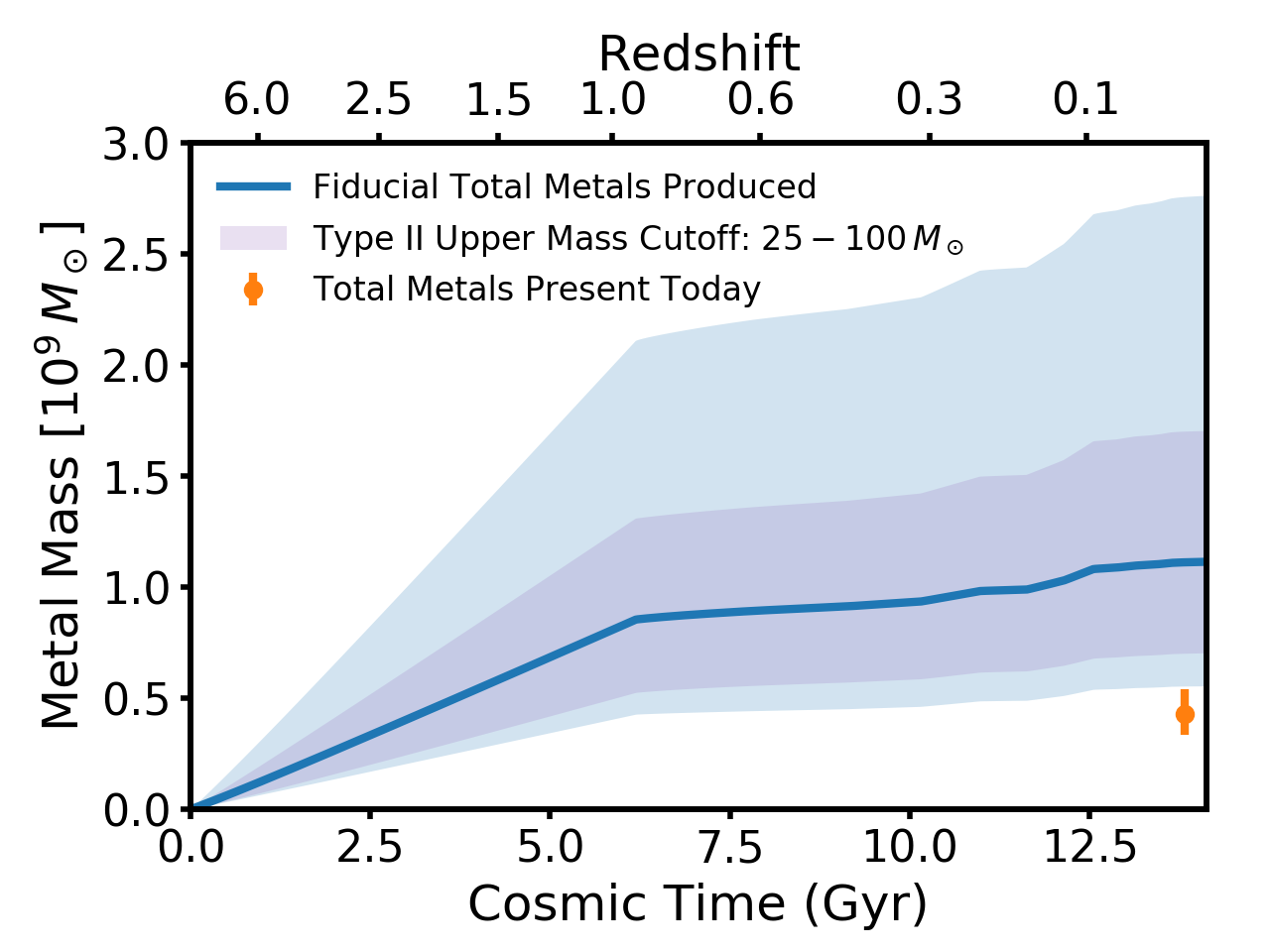}
\endminipage\hfill
\minipage{0.45\textwidth}
  \includegraphics[width=\linewidth]{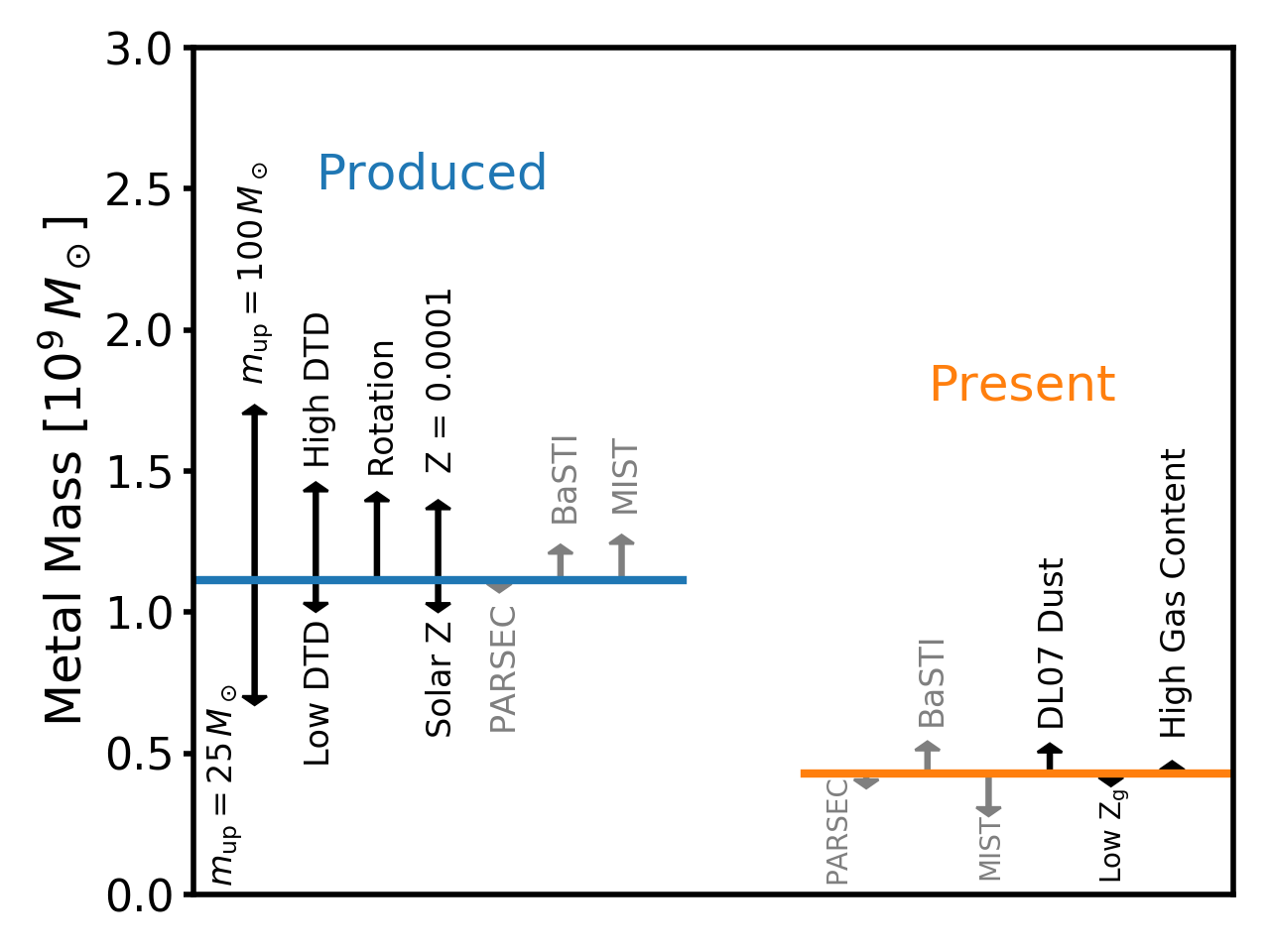}
\endminipage\hfill
\caption{\textsc{\textbf{Metal mass is missing from the M31 disk.}} Left: comparing the total produced and present metal mass within the PHAT footprint. The solid blue line shows the cumulative metal mass produced by all nucleosynthetic sources in the fiducial model up to the present day. The shaded blue region shows the allowed range of metal production within our conservative systematic uncertainty budget, and the shaded purple range illustrates that about half of the total uncertainty is due to the unknown $m_\mathrm{up}$, the highest mass star that explodes as a Type II SN and returns metals to the ISM. We compare the total produced metal mass to the fiducial present-day metal mass, shown as the orange point with an error bar to illustrate the allowed range within systematic uncertainties. Right: visualization of how various model parameters contribute to the systematic uncertainty budget. Arrows indicate the direction and magnitude of the change in total produced or present metal mass if a given extreme parameter choice is made instead of the fiducial model parameter (see Section~\ref{sec:metal_model} for details). The choice of stellar evolutionary tracks (and the solar abundance set adopted by each) affects both the produced and present metal mass. No combination of model parameters can result in more metal mass present in the PHAT footprint in M31 than was produced there, requiring that $f_\mathrm{retained} < 1$. Furthermore, the systematic uncertainty budget allows more freedom for higher metal production per mass of stars formed, and therefore lower $f_\mathrm{retained}$ than we find for our fiducial model.
\label{fig:metal_retention_total}}
\end{figure*}

We integrate the fiducial metal production histories over time to find the total metal mass that was produced in each SFH pixel, and then take the sum over all regions to calculate the metal mass that was produced across the PHAT footprint to find $1.1\times10^9\,M_\odot$ (with a possible range of $5.5\times10^{8}\,M_\odot - 2.8\times10^{9}\,M_\odot$ within conservative systematic uncertainties; Section~\ref{sec:metal_history}). We compare this to the total metal mass currently present in stars, gas, and dust, $4.3\times10^8\,M_\odot$ (with a possible range of $3.57-5.16\times10^8\,M_\odot$; Section~\ref{sec:metals_present}). The total missing metal mass from the PHAT region is the difference between the produced and present metal mass: $6.9\times10^8\,M_\odot$,  which is 61.5\% of the total metal production. Therefore, we find $f_\mathrm{retained}=38.5\%$ integrated over the PHAT footprint for the fiducial model. Within our conservative systematic uncertainty budget, we find allowed metal retention fractions $12.1 \% < f_\mathrm{retained}<97.4\%$, corresponding to a metal mass missing from the PHAT area of between $9.7\times10^7\,M_\odot$ and $6.3\times10^9\,M_\odot$.

We illustrate the strong constraint that metal mass is missing in Figure~\ref{fig:metal_retention_total}. The left panel shows the cumulative metal mass production history in the PHAT footprint as the solid blue line; the value at the oldest cosmic time is the total metal mass produced to the present day. The systematic uncertainty budget for the metal production model is shown as the blue shaded region, and the purple shaded region illustrates the large contribution to this budget of the unknown upper mass cutoff ($m_\mathrm{up}$) of stars that explode as Type II SNe and return metal mass to the ISM. The orange point shows the present-day metal content of the same area, with an error bar to illustrate the minimum and maximum possible metal mass. The range of possible present metal mass is derived from the $\sim20\%$ random errors in the SFH and systematic uncertainties in the gas-phase metallicity, hydrogen mass, and dust content. The same fiducial Padova SFH is used in all calculations in this plot to ensure a fair comparison. There is no overlap between the uncertainty in the cumulative metal mass produced to the present and the uncertainty in the current metal mass content, so even within a very large systematic uncertainty budget, we can state confidently that $f_\mathrm{retained} < 1$.

The right panel of Figure~\ref{fig:metal_retention_total} illustrates the effect of each uncertain model ingredient in our calculation of $f_\mathrm{retained}$, which were discussed in Sections~\ref{sec:metals_present_all} and ~\ref{sec:metal_model} above. The blue and orange lines show the total metal mass produced and present, respectively, for our fiducial calculations. The main uncertainties in the metal production model are the upper mass cutoff, $m_\mathrm{up}$, the effect of stellar rotation on Type II SN yields, the poorly constrained Type Ia SN DTD (time to first supernova explosion, slope, and normalization), and the metallicities of the progenitor stars (i.e., uncertainties in the stellar enrichment history). The choice of stellar evolutionary tracks in the SFH derivations affect both produced and present metal mass, but cannot drive the total produced metal mass any lower; our fiducial Padova SFHs therefore give a conservatively high $f_\mathrm{retained}$. The dominant uncertainty in the present metal mass is the choice of stellar evolutionary tracks (and the adopted solar abundance sets for each) because most present metal mass is in the stellar component. Although the gas-phase metallicity ($Z_\mathrm{g}$), hydrogen content, and normalization of the \citet{draine07} dust models are all uncertain, they cannot change the present-day metal budget much. 

A final systematic uncertainty that is not illustrated in Figure~\ref{fig:metal_retention_total} is the normalization of the SFHs and $M_\star$. This quantity is inherently uncertain at the factor of $\sim$2 level \citep[e.g.,][]{conroy13, courteau14} due to the unconstrained contribution of faint low-mass stars to the total $M_\star$ of a galaxy. However, an incorrect normalization of the SFHs would bias the produced metals and present metals in stars in the same direction (both either too high or too low). Therefore, a different normalization would change our $f_\mathrm{retained}$ only slightly, but could change the missing metal mass substantially. As an example, if we reduce the SFH normalization by a factor of $\frac{1}{2}$ compared to our fiducial model, we find $f_\mathrm{retained}=41.8\%$ within the PHAT footprint, but that the total missing metal mass is reduced to $3.2\times10^8M_\odot$. $f_\mathrm{retained}$ increases slightly because fewer metals are in stars, so the metals in the neutral ISM and dust become more important, while the missing metal mass is more strongly affected, decreasing by 53\% relative to the fiducial calculation. 

The takeaway from this discussion of systematic uncertainties is that our claim that metals are missing from the PHAT footprint is very secure. To obtain a higher $f_\mathrm{retained}$, model uncertainties would have to conspire to yield lower metal production and higher present-day metal mass; e.g., a very low $m_\mathrm{up}$ coupled with the highest possible stellar metal content. Our fiducial $f_\mathrm{retained}$ of 38.5\% is actually conservatively high; it is clear from Figure~\ref{fig:metal_retention_total} that there is more freedom for a higher produced metal mass and therefore lower $f_\mathrm{retained}$. 

Using our calculations of the total present and produced metal mass in the entire M31 disk (assuming azimuthal symmetry) presented in Sections~\ref{sec:metals_present} and ~\ref{sec:metal_history} above, we calculate that $1.8\times10^9 \, M_\odot$ of metals are missing within $r < 19\,\mathrm{kpc}$. Within our conservative systematic uncertainty budget, the total metal mass missing from the M31 disk could range from $1.9\times10^7\,M_\odot$ to $6.4\times10^9\,M_\odot$.  The metal retention fraction for the entire M31 disk is 38.4\% in the fiducial model, slightly lower than that in the PHAT area because regions at smaller radii contribute more; here, metals are retained less efficiently (see Section~\ref{sec:spatial_variation} below). 

For comparison, \citet{peeples14} find that galaxies with $M_\star \sim 10^{11}\,M_\odot$ have retained about 25\% of their metals, which is lower than the 38.3\% that we find for M31. Relative to our calculations, \citet{peeples14} find 55\% more present metals, but 115\% more metal production. These differences together predict that a larger fraction of metal mass was lost. A value of $f_\mathrm{retained}=25\%$ is well within our systematic uncertainty budget, above our lower limit of 12.1\%, so the discrepancy can be accounted for by different modeling choices. The relatively low $f_\mathrm{retained}$ found for M31 in both our study and in \citet{peeples14} implies that even high-mass galaxies have experienced significant mass loss; we consider this point quantitatively in Section~\ref{sec:outflows} below.

We may ask whether hierarchical accretion would affect our calculation of integrated $f_\mathrm{retained}$. When a large galaxy like M31 consumes a smaller satellite galaxy, essentially all of the stars from the satellite end up in the central galaxy, but some gas might be stripped during infall. Because we expect lower-mass galaxies to lose a greater fraction of their gas due to their shallow potential wells \citep[e.g.,][]{tremonti04, peeples11}, gas and metals are probably preferentially lost from the smaller progenitor galaxy before the merger even occurs. The removal of a greater fraction of gas (and the metals harbored in that gas) from the smaller satellite galaxy would drive the measured $f_\mathrm{retained}$ in the central galaxy downward.

However, this effect is probably small. If the accreted galaxy is much smaller than the central (1:$\gtrsim$10), then the accreted stars add less than 10\% to the expected metal production. Even in the extreme case where all of the gas and metals produced by those stars were removed from the satellite and never made it into the central galaxy $f_\mathrm{retained}$ would be biased low by less than 10\% (i.e., $f_\mathrm{retained}$ would decrease from 38.4\% to 34.8\%), which is well within the systematic uncertainty budget presented in Figure~\ref{fig:metal_retention_total}. If the accreted galaxy is larger, then it is less likely to lose significant gas and metal mass because its gravitational potential is larger. Therefore, we expect less dilution of the measured $f_\mathrm{retained}$ for more massive accreted satellites.

A competing effect is the accretion of pre-enriched material from the IGM, or via winds from external galaxies that are never accreted onto the central galaxy. These processes will add metal mass that was not produced by stars that currently reside in M31, resulting in a higher observed $f_\mathrm{retained}$. The total metal mass accreted in this manner is not expected to be high, but may at least in part balance the tendency of hierarchical accretion to drive $f_\mathrm{retained}$ slightly lower than its true value. Overall, the uncertainty due to hierarchical merging and accretion of pre-enriched gas is far smaller than our systematic uncertainty budget for $f_\mathrm{retained}$, so these effects will not bias our measurement of integrated $f_\mathrm{retained}$ in M31.

\subsection{Spatial Variation in Metal Retention\label{sec:spatial_variation}}

\begin{figure*}[!ht]
\minipage{0.45\textwidth}
  \includegraphics[width=\linewidth]{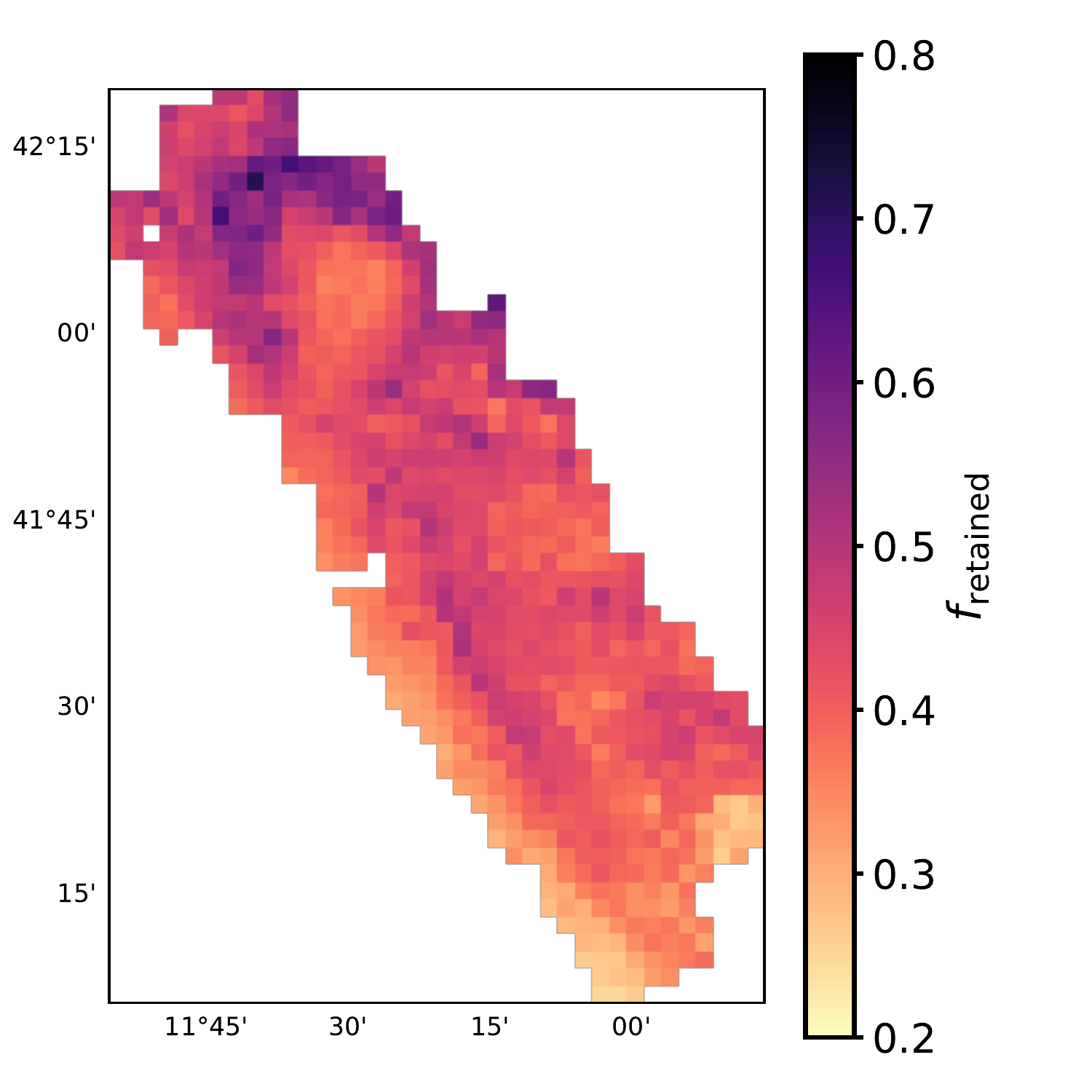}
\endminipage\hfill
\minipage{0.55\textwidth}
  \includegraphics[width=\linewidth]{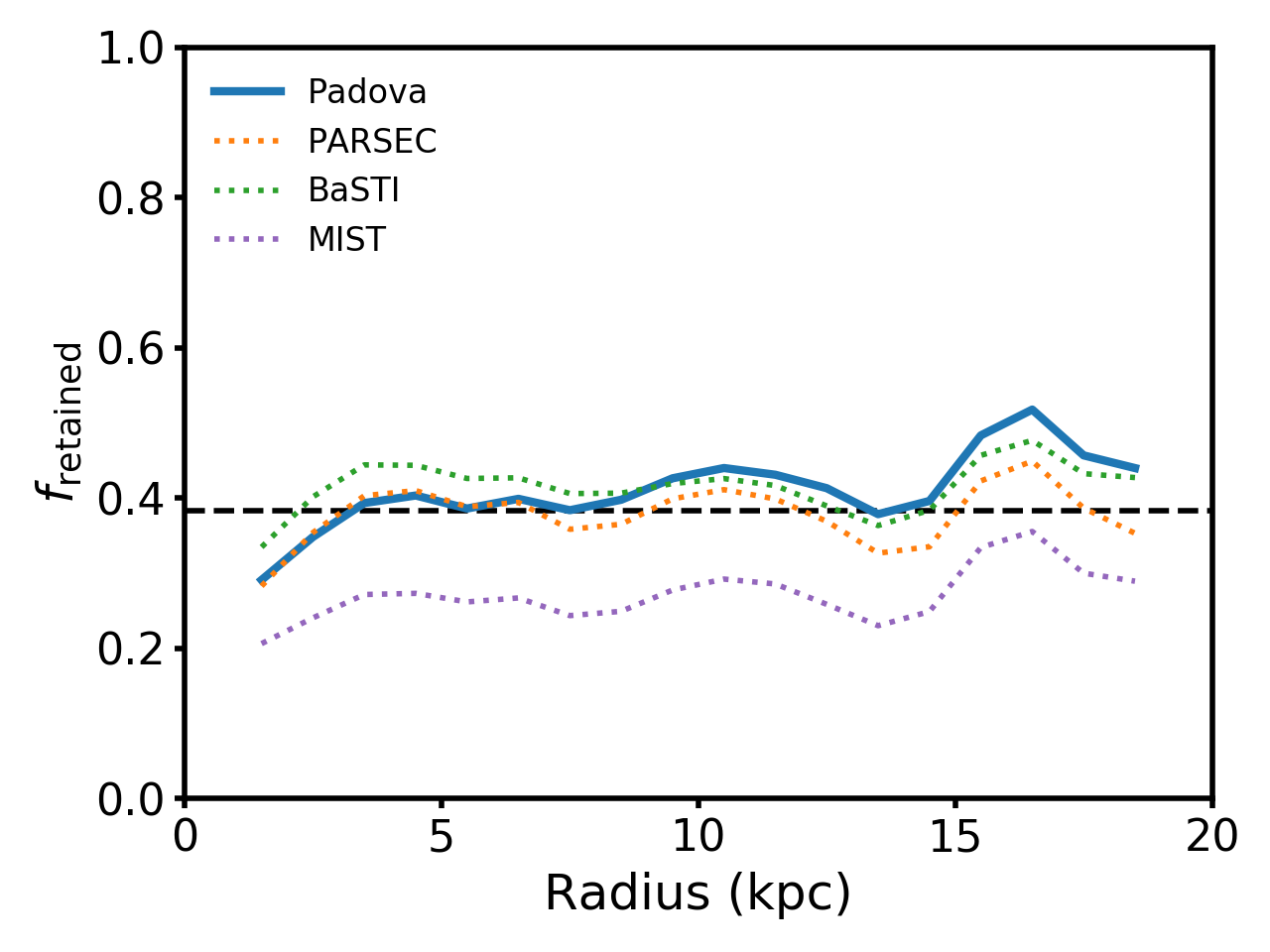}
\endminipage\hfill
\caption{\textsc{\textbf{The spatial variation in metal retention.}} Left: map of the metal retention fraction $f_\mathrm{retained}$ for the fiducial model within the PHAT footprint. In all SFH pixels, more metal mass was produced than is currently present, but gas-rich regions, tracing the star-forming rings, have higher $f_\mathrm{retained}$. Right: $f_\mathrm{retained}$, azimuthally averaged in $1 \mathrm{kpc}$ wide annuli, plotted against radius. The solid blue line shows $f_\mathrm{retained}$ obtained from the fiducial Padova SFHs, metal production model, and present-day metal calculations. The dashed black line shows the integrated $f_\mathrm{retained}$ for the entire PHAT area (38.4\%) for the fiducial model. The dotted lines show the variation in the radial $f_\mathrm{retained}$ profile that is due to the choice of stellar evolutionary tracks used in deriving the SFH. The SFHs and enrichment histories that were derived using each set of stellar evolutionary tracks (PARSEC in orange, BaSTI in green, and MIST in purple) were used to calculate the produced metals (with all other parameters fixed to fiducial values) and present metals in stars. There is little variation in the slope of the radial gradient in $f_\mathrm{retained}$ due to the choice of stellar evolutionary tracks.  
\label{fig_metal_retention_spatial}}
\end{figure*}

Here, we explore the spatial variation in $f_\mathrm{retained}$ within M31 and the sensitivity of the radial metal retention profile to the choice of stellar evolutionary models and to the redistribution of stars within the disk. The left panel of Figure~\ref{fig_metal_retention_spatial} shows a map of the PHAT region color-coded by $f_\mathrm{retained}$ for the fiducial model, defined as as the ratio of present to produced metal mass surface density in each SFH pixel. The center of M31 lies near the bottom right corner of this plot, so regions at the largest radii are found in the top left corner. From this map, it is obvious that the metal retention is not uniform across the disk. The highest metal retention fractions are found in regions with the highest gas content, tracing the star-forming rings (purple areas). For our fiducial calculation, no SFH pixels contain more metal mass today than was produced there (i.e., $f_\mathrm{retained} < 1$ everywhere).

The right panel of Figure~\ref{fig_metal_retention_spatial} shows the metal retention fraction averaged in 1 kpc wide annuli. The solid blue line shows the radial profile of $f_\mathrm{retained}$ for our fiducial Padova SFHs. The dashed black line shows the total $f_\mathrm{retained}$ for the PHAT footprint, 38.5\%. The dotted lines show the results when the SFHs and enrichment histories derived using different stellar evolutionary tracks (and the solar abundances adopted by each model set) are used to calculate the produced and present metal mass. The results from PARSEC are shown in orange, BaSTI in green, and MIST in purple. 

Figure~\ref{fig_metal_retention_spatial} shows a weak radial gradient in $f_\mathrm{retained}$, such that a larger fraction of produced metal mass is missing in the inner regions than in the outer regions. As discussed in Section~\ref{sec:metals_present} above, over 90\% of the metal mass that is present is in the stellar component. The fraction of produced metal mass retained in stars is roughly constant with radius, so this weak gradient is due to the increased contribution of neutral gas and dust to the present-day metal budget at larger radii. The metals in gas and dust are found predominantly in the gas-rich rings near 10 and 17 kpc, causing $f_\mathrm{retained}$ to increase by about 10\% in the outskirts relative to the gas-poor central regions.

There is little variation in the slope of the weak radial gradient among the four $f_\mathrm{retained}$ profiles obtained from the different stellar evolutionary tracks. The fiducial Padova SFHs and the MIST SFHs both result in slightly steeper gradients than the BaSTI or PARSEC models. The uncertain ingredients in the metal production model only affect the normalization of the $f_\mathrm{retained}$ radial profile, not the slope.
The total $f_\mathrm{retained}$ is similar among the Padova, PARSEC, and BaSTI models, but is lower (i.e., more metal loss) for the MIST models, which adopt a lower absolute value for solar metallicity (see Section~\ref{sec:sfhs}). The [M/H] vs. time enforced in the CMD modeling produces a lower absolute value for the metallicity of the stars formed in the MIST models, driving the metal mass production higher (see the right panel of Figure~\ref{fig:metal_retention_total}).

The CMD-based SFHs are only sensitive to the present-day location of stars, not where they formed initially. Therefore, it is likely that both the metal production histories and present-day metal mass that we calculate at a given radius include contributions from star formation that actually occurred at a range of radii. 

The stars in the M31 disk have likely migrated radially away from their birth locations over many gigayear timescales. Generally, stars born at small radii experience a net outward migration, such that at a fixed radius in the disk, a greater fraction of stars currently present were born interior to that radius than exterior to it \citep{roskar08, bird12}. Furthermore, multiple lines of evidence suggest that M31 experienced a merger or interaction in the recent past \citep[e.g.,][]{dsouza18, hammer18}, including its dynamically hot disk \citep{dorman15}, the wealth of structure in its stellar halo \citep{komiyama18}, and the global burst of star formation $\sim2-3\,\mathrm{Gyr}$ ago \citep{williams15}. Such an event would have further disrupted stellar orbits.

The redistribution of stars (and the metals they contain) within the disk of M31 does not affect our integrated $f_\mathrm{retained}$, but likely plays a role in shaping the shallow radial gradient in $f_\mathrm{retained}$ that we measure. Stars are the dominant present-day metal reservoir at all radii in the M31 disk, and we find little variation the fraction of produced metals retained in stars as a function of radius. It remains entirely possible that there was a stronger radial gradient in metal retention that has been washed out by the redistribution of stars. 


\section{Discussion: Metal Ejection and Redistribution\label{sec:discussion}}
We have compared the total metal mass produced in the M31 disk ($r < 19\,\mathrm{kpc}$) to a census of metals present today, and found that just 38\% of metal mass has been retained in the disk for our fiducial model (Section~\ref{sec:missing_mass}). Furthermore, we have demonstrated that no combination of model parameters can conspire to account for all produced metal mass in the M31 disk today, so $f_\mathrm{retained}$ must be less than unity. Here, we use this result to constrain the lifetime-averaged mass-loading of outflows and the expected metal content of M31's CGM. We then use the spatially resolved metal production histories (Section~\ref{sec:metal_history}) to show that metals have likely been transported from the central regions outward within the disk over the past 1.5 Gyr.

\subsection{Lifetime-averaged Outflow Properties\label{sec:outflows}}

The finding that even massive galaxies have lost metals over their lifetimes is consistent across several studies \citep{zahid12b, peeples14, belfiore16a}. The implication is that that metals have been entrained in gaseous outflows, removed from central galaxies, and deposited in their CGM and/or the IGM. Constraining past outflows from galaxies is a key goal of galaxy evolution studies; in particular, knowledge of the total gas mass lost from galaxies and its ultimate fate, as a function of galaxy stellar mass, would provide valuable insight into the operation of the cosmic baryon cycle. Here, we use our fiducial calculation of the lifetime-integrated metal loss from M31 to constrain the time-averaged metal and baryon content of the outflows that carried those metals out of the disk.

Previous studies have inferred the total mass and metal content of outflows from observations \citep[e.g.,][]{weiner09, rubin14, chisholm18}. However, such observational constraints are only sensitive to the metal-enriched material in certain phases of the wind, typically cool, dense clouds entrained in the wind (as traced by, e.g., Mg \textsc{ii} absorption). Inferences of the \textit{total} metal and mass outflow from observations therefore require highly uncertain corrections for the ionization structure, geometry, and metallicity of the wind. Furthermore, some of the material entrained in ongoing outflows may quickly fall back onto the galaxy, so the outflow rates inferred from observations include some material that would not be considered ``lost" from the galaxy over long timescales. 

In contrast to observational measurements, which are inherently only sensitive to a subset of the outflowing material, theoretical chemical evolution models or hydrodynamical simulations typically include all metals and baryons entrained in the wind, across all phases, in defining the metal and mass loading of winds \citep[e.g.,][]{muratov15, belfiore16a}. Our fiducial calculation of the total metal mass lost from M31's disk over its lifetime enables us to constrain the lifetime-averaged metal content of outflows from M31, including metals lost in all wind phases. The calculations that follow are therefore directly comparable to mass-loading factors in chemical evolution models or galaxy formation simulations, but are fundamentally different quantities than those measured from observations of particular wind phases.

First, we consider the metal content of past outflows from M31. We adopt the \citet{peeples11} definition of metal expulsion efficiency, $\zeta_\mathrm{wind}$, which describes the rate at which metal mass is ejected from a galaxy:

\begin{equation}
\zeta_\mathrm{wind} = \frac{Z_\mathrm{wind}}{Z_\mathrm{ISM}} \, \frac{\dot{M}_\mathrm{wind}}{\mathrm{SFR}} = \frac{Z_\mathrm{wind}}{Z_\mathrm{ISM}} \, \eta_\mathrm{wind},
\end{equation}
where $Z_\mathrm{wind}$ and $Z_\mathrm{ISM}$ are the outflow and ISM metallicities, $\dot{M}_\mathrm{wind}$ is the total mass outflow rate, and $\eta_\mathrm{wind}$, the ratio of mass outflow rate to star formation rate, is called the baryonic mass-loading factor. Again, both $\zeta_\mathrm{wind}$ and $\eta_\mathrm{wind}$ include all metals and baryons in all phases of the wind. Physically, $\zeta_\mathrm{wind}$ can be thought of as the ratio of the rate of metal ejection to the rate at which metals are locked into stars. Under the assumption that the metal expulsion efficiency is constant in time, we can approximate the lifetime-averaged $\left<\zeta_\mathrm{wind}\right>$ for M31 as

\begin{multline}
\left<\zeta_\mathrm{wind}\right> = \frac{\int \frac{Z_\mathrm{wind}\dot{M}_\mathrm{wind}}{Z_\mathrm{ISM}\mathrm{SFR} }\mathrm{d}t }{\Delta t} = \frac{Z_\mathrm{wind}\dot{M}_\mathrm{wind} \Delta t}{Z_\mathrm{ISM}\mathrm{SFR} \Delta t} \\
 = \frac{M_Z^\mathrm{missing}}{M_Z^\mathrm{formed\,stars}}.
\end{multline}
Here, $\Delta t$ is the age of the universe, $M_Z^\mathrm{missing}$ is the total missing metal mass, and $M_Z^\mathrm{formed\,stars}$ is the total metal mass that was ever incorporated into stars formed in M31, and so is higher than the current metal mass in stars by a factor of $1 / (1-R) \simeq 1.7$.

We find a lifetime-averaged $\left<\zeta_\mathrm{wind}\right>=1.05$ in M31 for our fiducial model. This metal expulsion efficiency is consistent with theoretical expectations for an M31-mass galaxy based on analytic modeling of the observed mass-metallicity relation \citep[$\zeta_\mathrm{wind}\sim0.2-1.6$, depending on the choice of metallicity calibration;][]{peeples11}. Recent measurements of the total metal expulsion efficiency in metal-enriched outflows from $M_\star\sim10^{10-11} \, M_\odot$ galaxies by \citet{chisholm18} that include careful corrections for the ionization structure and geometric effects also agree with the \citet{peeples11} expectation. Adopting our bounding calculations, which produce the minimum and maximum missing metal mass, we find that $0.02 < \left<\zeta_\mathrm{wind}\right> < 4.6$ within conservative systematic uncertainties.

Next, we consider implications for the mass-loading factor, $\eta_\mathrm{wind}$, which is of particular interest for understanding the cosmic baryon cycle. Different feedback models used in galaxy formation simulations result in mass-loading factors that differ in both normalization (i.e., the total amount of gas that is removed from galaxies, including all wind phases) and scaling with stellar mass \citep[e.g.,][]{muratov15, christensen18}. Unfortunately, the degeneracy between $Z_\mathrm{wind}$ and $\dot{M}_\mathrm{wind}$ limits our ability to constrain $\eta_\mathrm{wind}$ from either observations of metal-enriched winds or measurements of total metal loss. The metal mass lost over M31's lifetime could have been expelled in very metal rich winds, or in winds with metallicity comparable to that of the ambient ISM; these scenarios would require relatively low or high mass-loading factors, respectively.

We can place an upper limit on the lifetime-averaged mass loading of outflows from M31 by assuming $Z_\mathrm{wind} \gtrsim Z_\mathrm{ism}$, because there is no physical reason that outflows should be metal-poor compared to the ambient gas. Outflows may be enriched relative to the ISM metallicity because the gas entrained in the outflows is likely to be dominated by supernova ejecta. In this case, $\eta_\mathrm{wind} \lesssim1.05$, meaning that M31 could have ejected as much gas as has gone into forming stars (but no more). A value of $\eta_\mathrm{wind} \sim 1$ is consistent with the findings of \citet{belfiore16a}, who assumed $Z_\mathrm{wind} = Z_\mathrm{ism}$ in their chemical evolution modeling of NGC 628, a galaxy whose stellar mass is $\sim10\times$ lower than that of M31, and found that their spatially resolved metal budget is well described by a constant $\eta_\mathrm{wind} = 1$.

If $Z_\mathrm{wind}/ Z_\mathrm{ISM} > 1$, then a lower value of $\eta_\mathrm{wind}$ is required to produce the same metal expulsion efficiency. Physical expectations and observational constraints suggest that outflows are typically metal enriched relative to the ambient ISM \citep[e.g.,][]{martin02, chisholm18}. State-of-the-art feedback models implemented in recent zoom-in simulations of galaxies \citep[e.g.,][]{angles-alcazar17, christensen18} generally result in outflows that are enriched relative to the ISM, although $Z_\mathrm{wind}/ Z_\mathrm{ISM}$ varies in simulations that employ different feedback models. Taken together, these results suggest that outflows from M31 were metal enriched relative to its ISM, and that the total gas mass ejected from M31's disk was lower than the total stellar mass formed.

\subsection{The Metal Content of the M31 CGM\label{sec:cgm}}

Galaxies are now known to harbor massive gaseous halos that mediate the exchange of baryons and metals between galaxies and the IGM \citep{tumlinson17}. The CGM of $L*$ galaxies are ubiquitously metal enriched \citep[e.g.,][]{lehner13, werk14, prochaska17}, likely harboring a large fraction of metals that have been ejected from their central galaxies via outflows, in addition to metals lost from external satellite galaxies. The standard method of studying the CGM is to measure the absorption by CGM material in the spectra of quasars whose sightlines pass through the CGM of foreground galaxies. For the vast majority of CGM detections, we only sample a single line of sight, and so it remains unclear how the properties of the CGM vary within a single halo (vs.\ halo-to-halo variability).

M31 is a unique target for characterizing the CGM of an $L*$ galaxy using many quasar sightlines because it is nearby and subtends a large area on the sky. \citet{lehner15} have used this technique to place a lower limit on the CGM metal content of $\gtrsim 2\times10^6\,M_\odot$ (within the inner $50\,\mathrm{kpc}$) using detections of $\mathrm{Si\,\textsc{ii}}$, $\mathrm{Si\,\textsc{iii}}$, and $\mathrm{Si\,\textsc{iv}}$ absorption along several quasar sightlines through the M31 CGM. Extrapolating out to the virial radius, these authors calculate a minimum of $1.4\times10^7\,M_\odot$ of metals in M31's CGM. However, the metal ions that produced the absorption features in their study only trace the cool-warm CGM material ($T \sim 10^{4-5.5}\,\mathrm{K}$). Unfortunately, for a massive spiral galaxy like M31, most of the CGM material is expected to reside in a hot corona \citep[e.g.,][]{white78, birnboim03}. 

As reported in Section~\ref{sec:missing_mass} above, $1.8\times10^9 \, M_\odot$ of metal mass is missing from the M31 disk (at $r < 19\,\mathrm{kpc}$) according to our fiducial model (with a conservative range between $1.9\times10^7\,M_\odot$ and $6.4\times10^9\,M_\odot$), over a factor of 100 greater than the lower limit placed on metal mass in the cool CGM by \citet{lehner15}. In the scenario where all ejected metals and baryons are now harbored in the CGM, the metal mass missing from M31 can be viewed as a predicted lower limit on the expected metal content of the M31 CGM, given that additional metals could have been ejected from the outer disk of M31 (at $r > 19\,\mathrm{kpc}$), or deposited from pre-enriched material accreted from the IGM. 

Next, we ask if it is plausible for the metal mass missing from the disk of M31 to be harbored in its CGM. We combine our fiducial calculation of missing metal mass with an estimate of the total gas content of M31's halo to calculate the implied metallicity of the CGM if all missing metals are retained within the halo. We then compare to the observed distribution of CGM metallicities for $L*$ galaxies and ask whether required metallicity of M31's CGM is consistent with observations. If the implied CGM metallicity is much higher than typical CGM metallicities for similar-mass galaxies, it would suggest that metals may have been lost beyond M31's halo into the IGM. This calculation includes metals and gas in all phases of the CGM. 

We adopt a dark matter halo mass for M31 of $M_\mathrm{halo}=1.0\times 10^{12}\,M_\odot$ \citep{tamm12}, although we note that this quantity is uncertain at the factor of $\sim2$ level \citep[e.g.,][]{penarrubia14}. We assume that the halo contains enough hydrogen for its baryon fraction to be equal to the cosmic $f_\mathrm{baryon}=0.164$ \citep{hinshaw13}, and after accounting for the stellar mass, ISM mass, and contribution of helium ($M_\mathrm{He} = 0.36 M_\mathrm{H}$) to the CGM mass, we find a total mass of hydrogen in all phases of the CGM of $3.9\times10^{10} \, M_\odot$. Given this hydrogen mass, our fiducial missing metal mass of $1.8\times10^9 \, M_\odot$ requires a CGM metallicity of $2.4\,Z_\odot$ (adopting $Z_\odot = 0.019$), or $\mathrm{[Z/H]} = 0.4$, if all metals are retained in M31's halo. 

\citet{prochaska17} measured CGM metallicities along individual sightlines through $L*$ galaxy halos at $z\sim0.2$, finding a median value of $[\mathrm{M/H}]=-0.5$ and a 95\% confidence interval between $-1.7 \lesssim \mathrm{[Z/H]} \lesssim 0.7$. The wide range reflects a combination of variation in metallicity \textit{between} halos and variation \textit{within} individual halos because the metallicity along a given sightline is not necessarily equal to the average metallicity of that halo. For our fiducial missing metal calculation, the average M31 CGM metallicity is consistent with, though at the high end of, the \citet{prochaska17} distribution. Considering our minimum and maximum constraints on missing metal mass, we find a possible range of metallicities for M31's CGM between $-1.6 \lesssim \mathrm{[Z/H]} \lesssim 1.0$. This range is wider than the observed distribution of CGM metallicities, but we emphasize that our bounding metal loss calculations adopt extreme assumptions. 

The consistency between the CGM metallicity implied by our fiducial missing metal mass and the observed range of CGM metallicities for $L*$ galaxies suggests that all metals missing from M31's disk can plausibly remain in its halo. Two uncertain factors could drive our predicted CGM metallicity even higher: (1) if the $f_\mathrm{baryon}$ in galaxy halos is lower than the cosmic value \citep[e.g.,][]{anderson-m10, ford14}, our estimated hydrogen mass in the halo would decrease; and (2) if metal-enriched material from the IGM or winds from satellite galaxies \citep[e.g.,][]{oppenheimer06, angles-alcazar17} has accreted onto M31's halo, the metal mass in the halo would increase. If future characterization of the M31 CGM determines that its metallicity is much lower than the predicted $\mathrm{[Z/H]} = 0.4$ for our fiducial metal loss calculation, this would suggest that metal-enriched material has been lost from the halo to the IGM. 

\subsection{Recent Metal Transport within the Disk of M31\label{sec:transport}}

\begin{figure}[!t]
  \includegraphics[width=\linewidth]{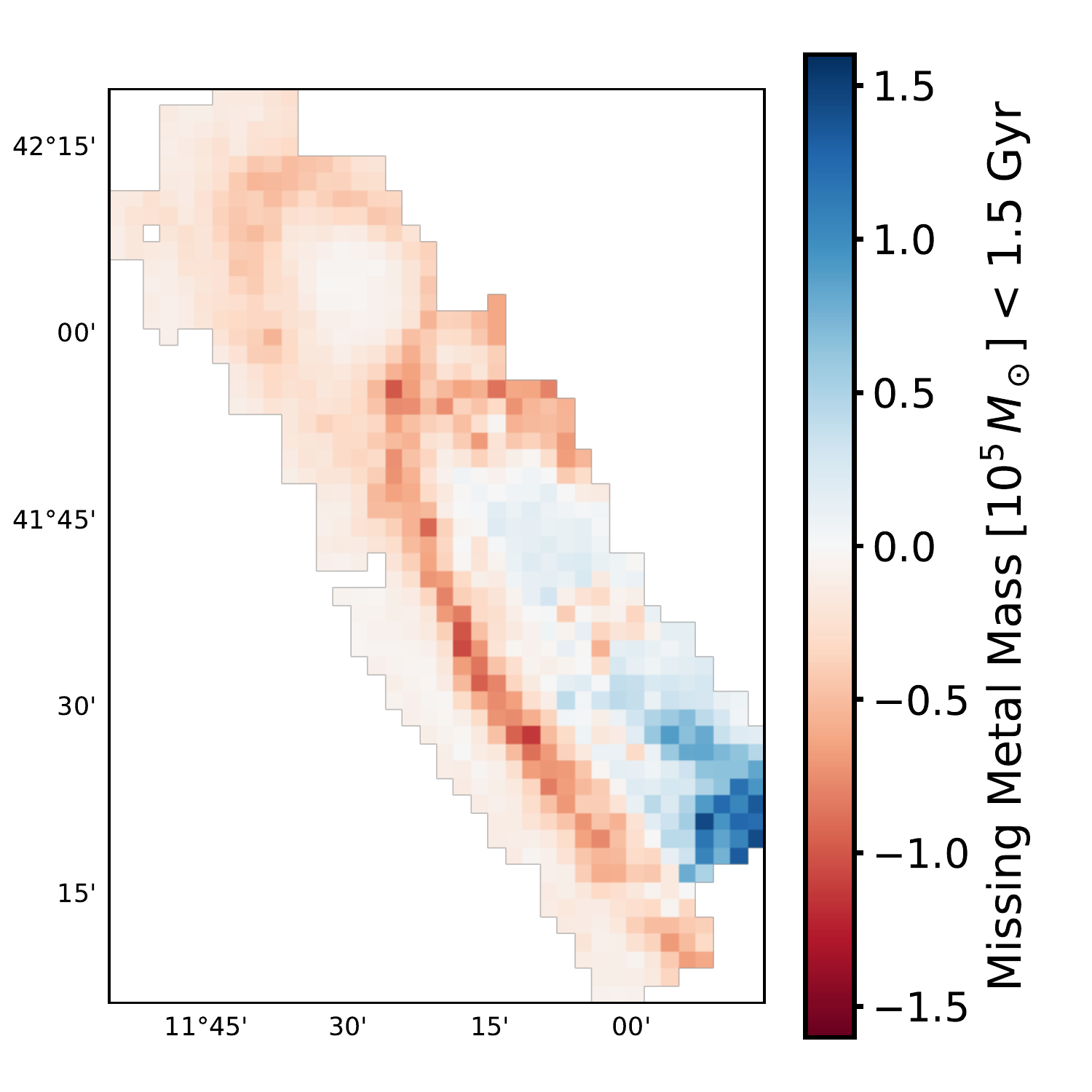}
\caption{\textsc{\textbf{Recently produced metals have been transported out of the central disk.}} Map of the difference between the total metal mass produced in each SFH pixel in the past 1.5 Gyr and the metal mass present there that could have been formed since that time, calculated as the sum of metal mass in the neutral ISM, in dust, and in stars that formed in the past 1.5 Gyr. SFH pixels colored blue (near the center of the disk) have produced more metals since 1.5 Gyr ago than can be accounted for in those regions, while red SFH pixels harbor an excess of present-day metal mass that could have been produced recently. This map shows that although no net metal loss from the PHAT footprint is required in the past 1.5 Gyr, recently produced metal mass is preferentially missing from the central regions and may have been transported outward in the disk.
\label{fig:recent_metals}}
\end{figure}

In this section, we leverage our knowledge of the metal production history and enrichment history of stars to calculate the lookback time before which outflows must have occurred. We cannot measure \textit{when} the metals were deposited into the neutral ISM and in the dust, but because we know the stellar enrichment history, we do know when metals were incorporated into the stars formed in a given age bin. To assess whether metals must have been removed from the PHAT footprint since a given lookback time, we calculate the difference between total metal mass formed since that time and the total metal mass that is present in the stars formed since then, in the neutral ISM, and in dust. If this difference is positive, then more metal mass was formed since that lookback time than could possibly have been incorporated into the neutral ISM, dust, and recently formed stars, and so metal mass must have been removed over that time interval.

We calculate the required metal mass loss since the beginning of each time bin, integrated over the entire PHAT footprint. By definition, the required metal loss since the beginning of the oldest time bin is equal to the total metal mass lost (Section~\ref{sec:missing_mass}). We find that the ``required metal loss'' is negative up to 1.5 Gyr ago, indicating that no net metal loss is actually required up to that lookback time. Interestingly, this lookback time is the beginning of the age bin immediately following the global burst of star formation. Increasing the lookback time to include the burst, we find that net metal mass loss becomes positive, and so metals must have been lost from the PHAT footprint since the beginning of the global burst of star formation $\sim2-3\,\mathrm{Gyr}$ ago.

Although no net metal loss is required from the M31 disk in the past 1.5 Gyr, we can perform the same calculation within each SFH pixel and ask whether metal mass must have been removed from some regions, i.e., whether metals have been redistributed within the disk. Figure~\ref{fig:recent_metals} shows a map of the PHAT footprint color-coded by the difference between total metal mass produced in the past 1.5 Gyr and the total metal mass present in the neutral ISM, dust, and stars formed in the past 1.5 Gyr.  SFH pixels shown in red contain an excess of metal mass that could have been deposited over that timescale over the metal mass that has formed over the same timescale. The blue SFH pixels have formed more metal mass recently than can be accounted for in the neutral ISM, dust, and recently formed stars, and so metals must have been removed from these regions in the past 1.5 Gyr.

The pattern of net metal loss from the central regions and excess ``recently incorporated'' metal mass present in the outskirts over the expected recent metal production in those regions suggests that metal mass has been transported outward in the disk over the past 1.5 Gyr. Because we only consider metals produced since the global burst of star formation, which possibly coincided with a merger or interaction, we do not expect the stellar disk to have been disrupted since these relatively young stars formed and produced new metals. Stellar radial migration probably has occurred over this timescale, but because there is no radial variation in the fraction of produced metal mass that is retained in the stellar component, it should not affect this calculation. 

A total of $8.3\times10^5\,M_\odot$ of metal mass must have been moved outward in the disk over a typical distance of $\lesssim5\,\mathrm{kpc}$. This level of metal transport is consistent with turbulent mixing at the sound speed of the ISM, $c_s \sim 10 \, \mathrm{km \, s}^{-1}$ \citep[following][]{werk11}, corresponding to a maximum distance of $15\,\mathrm{kpc}$ traveled over 1.5 Gyr. It is possible that metals were instead mixed into a hot gaseous halo, and that some metal mass in the outer parts of the disk has been incorporated from wind recycling or enriched accretion from beyond $20\,\mathrm{kpc}$ \citep[e.g.][]{oppenheimer08}.


\section{Conclusions\label{sec:conclusions}}
\begin{enumerate}
\item We performed a census of the metal mass currently in the M31 disk. The stars harbor over 90\% of metals, but the fractional contribution of metals in the neutral ISM and dust is higher in gas-rich regions that trace the star-forming rings (Section~\ref{sec:metals_present}, Figure~\ref{fig:metals_present}).
\item We constructed a model of metal production by Type II SNe, AGB stars, and Type Ia SNe following a burst of star formation and bounded the systematic uncertainties in this model (Section~\ref{sec:metal_model}, Figure~\ref{fig:metal_model}). We convolved this model with the CMD-based spatially resolved SFHs derived from PHAT data to calculate the history of metal production in M31 (Section~\ref{sec:metal_history}, Figure~\ref{fig:metal_history}).
\item Integrated over the PHAT footprint, we calculated that 62\% of metal mass formed there is missing for our fiducial model. We show that $f_\mathrm{retained}<1$ for all possible model choices ($12.1 \% < f_\mathrm{retained}<97.4\%$), so metal loss is required even when all uncertain model parameters are chosen to favor metal retention. Assuming azimuthal symmetry to extend our calculations to the entire M31 disk, we found that $1.8\times10^9\,M_\odot$ of metal mass is missing from the M31 disk within $r < 19\,\mathrm{kpc}$ (with a possible range between $1.9\times10^7\,M_\odot$ and $6.4\times10^9\,M_\odot$; Section~\ref{sec:missing_mass}, Figure~\ref{fig:metal_retention_total}).
\item We show that there is little variation in the metal retention fraction with radius, athough $f_\mathrm{retained}$ does increase by about 10\% from the central regions to the most gas-rich regions at larger radii, creating a shallow positive radial gradient in $f_\mathrm{retained}$. The slope of the $f_\mathrm{retained}$ radial profile is insensitive to the choice of stellar evolutionary tracks used to derive the SFHs and enrichment histories (Section~\ref{sec:spatial_variation}, Figure~\ref{fig_metal_retention_spatial}).
\item From our fiducial calculation of lifetime metal mass loss from M31, we calculated a lifetime-averaged metal expulsion efficiency $\left<\zeta_\mathrm{wind}\right>=1.05$, with a possible range between $0.02 < \left<\zeta_\mathrm{wind}\right> < 4.6$. This places an upper limit on the mass-loading factor $\eta_\mathrm{wind} \lesssim 1.05$ for the fiducial model, implying that M31 could not have lost more gas mass over its lifetime than the total gas mass that went into star formation (Section~\ref{sec:outflows}). 
\item The missing metal mass we found for the fiducial model is over a factor of 100 higher than the available lower limit on the metal mass in M31's cool-warm CGM. The missing metal mass could be harbored in M31's CGM if the majority of those metals reside in a hot $T \gtrsim 10^{5.5}\,\mathrm{K}$ corona. We showed that the implied CGM metallicity of $[\mathrm{Z/H}] \sim 0.4$ for our fiducial calculation is within, but at the high end of, the range of CGM metallicities found for $L*$ galaxies (Section~\ref{sec:cgm}). 
\item We used the spatially resolved SFHs and fiducial metal production histories to show that no net metal loss from the M31 disk is required in the past 1.5 Gyr, but there must have been a net loss of metals produced in M31 before and during the global burst of star formation $2-3\,\mathrm{Gyr}$ ago. We found that metals produced in the past 1.5 Gyr are missing from the central regions, suggesting that material has either been ejected from the inner disk or transported radially outward over that time interval (Section~\ref{sec:transport}, Figure~\ref{fig:recent_metals}).
\end{enumerate}
\bigskip
\acknowledgements
We thank Andreas Schruba, Christa Gall, Molly Peeples, Gurtina Besla, Ekta Patel, Mary Putman, Charlie Conroy, Phil Hopkins, and Andy Dolphin for illuminating discussions and helpful feedback, and we also thank the referee for thorough and constructive comments that improved the clarity of this manuscript. O.G.T. is supported by an NSF Graduate Research Fellowship under grant DGE-1256082, and was supported in part by NSF IGERT grant DGE-1258485. J.W. acknowledges partial support from a 2018 Alfred P. Sloan Research Fellowship. 

This research has made use of NASA's Astrophysics Data System and the arXiv preprint server. This work was supported by the Space Telescope Science Institute through GO-12058. This work is based on observations made with the NASA/ESA \textit{Hubble Space Telescope}, obtained from the data archive at the Space Telescope Science Institute. STScI is operated by the Association of Universities for Research in Astronomy, Inc. under NASA contract NAS 5-26555. This research has made use of Montage. It is funded by the National Science Foundation under grant No. ACI-1440620, and was previously funded by the National Aeronautics and Space Administration's Earth Science Technology Office, Computation Technologies Project, under Cooperative Agreement Number NCC5-626 between NASA and the California Institute of Technology.

\software{iPython \citep{ipython}, Astropy \citep{astropy}, matplotlib \citep{matplotlib}, NumPy \citep{numpy}, HDF5 \citep{hdf5}, Montage \citep{berriman03, jacob10}}


\end{document}